\documentclass[12pt,aps,prd,tightenlines,groupedaddress,preprint,floatfix,nofootinbib]{revtex4}
\usepackage{amssymb,amsmath,graphicx,multirow}

\newcommand{\PRE}[1]{{#1}} 

\begin{document}

\preprint{UH-511-1178-11}

\title{\PRE{\vspace*{0.8in}}
Large Jet Multiplicities and New Physics at the LHC
\PRE{\vspace*{0.3in}}
}
\author{Joseph Bramante\footnote{E-mail address:  {\tt bramante@hawaii.edu}},
        Jason Kumar\footnote{E-mail address:  {\tt jkumar@hawaii.edu}},
      and Brooks Thomas\footnote{E-mail address:  {\tt thomasbd@phys.hawaii.edu}}}
\affiliation{Department of Physics, University of Hawaii, Honolulu, HI 96822 USA
\PRE{\vspace*{.5in}}
}

\begin{abstract}
\PRE{\vspace*{.3in}}
  A broad class of scenarios for new physics involving additional strongly-interacting
  fields generically predicts signatures at hadron colliders which consist solely of large
  numbers of jets and substantial missing transverse energy.
  In this work, we investigate the prospects for discovery in such
  scenarios using a search strategy in which jet multiplicity and missing transverse
  energy are employed as the primary criteria for distinguishing signal from
  background.  We examine the discovery reach this
  strategy affords in an example theory (a simplified supersymmetric model
  whose low-energy spectrum consists of a gluino, a light stop, and a light neutralino)
  and demonstrate that it frequently exceeds the reach obtained via other, 
  alternative strategies.
\end{abstract}


\maketitle


\newcommand{\newc}{\newcommand}
\newc{\gsim}{\lower.7ex\hbox{$\;\stackrel{\textstyle>}{\sim}\;$}}
\newc{\lsim}{\lower.7ex\hbox{$\;\stackrel{\textstyle<}{\sim}\;$}}

\def\vac#1{{\bf \{{#1}\}}}

\def\beq{\begin{equation}}
\def\eeq{\end{equation}}
\def\beqn{\begin{eqnarray}}
\def\eeqn{\end{eqnarray}}
\def\calM{{\cal M}}
\def\calV{{\cal V}}
\def\calF{{\cal F}}
\def\half{{\textstyle{1\over 2}}}
\def\quarter{{\textstyle{1\over 4}}}
\def\ie{{\it i.e.}\/}
\def\eg{{\it e.g.}\/}
\def\etc{{\it etc}.\/}


\def\inbar{\,\vrule height1.5ex width.4pt depth0pt}
\def\IR{\relax{\rm I\kern-.18em R}}
 \font\cmss=cmss10 \font\cmsss=cmss10 at 7pt
\def\IQ{\relax{\rm I\kern-.18em Q}}
\def\IZ{\relax\ifmmode\mathchoice
 {\hbox{\cmss Z\kern-.4em Z}}{\hbox{\cmss Z\kern-.4em Z}}
 {\lower.9pt\hbox{\cmsss Z\kern-.4em Z}}
 {\lower1.2pt\hbox{\cmsss Z\kern-.4em Z}}\else{\cmss Z\kern-.4em Z}\fi}
\def\wtg{\widetilde{g}}
\def\wtq{\widetilde{q}}
\def\st1{\widetilde{t}_1}
\def\mst1{m_{\widetilde{t}_1}}
\def\Neut1{\widetilde{N}_1}
\def\N1{\widetilde{N}_1}
\def\Lint{\mathcal{L}_{\mathrm{int}}}

\newcommand{\Dsle}[1]{\hskip 0.09 cm \slash\hskip -0.28 cm #1}
\newcommand{\met}{{\Dsle E_T}}
\newcommand{\mht}{{\Dsle H_T}}
\newcommand{\Dslp}[1]{\slash\hskip -0.23 cm #1}
\newcommand{\mpt}{{\Dslp p_T}}
\newcommand{\mptvec}{{\Dslp \vec{p}_T}}
\newcommand{\bigDsle}[1]{\hskip 0.05 cm \slash\hskip -0.38 cm #1}
\newcommand{\bigmet}{{\bigDsle E_T}}

\newcommand{\gev}{{\rm GeV}}
\newcommand{\tev}{{\rm TeV}}
\newcommand{\ifb}{{\rm fb^{-1}}}
\newcommand{\ipb}{{\rm pb^{-1}}}
\newcommand{\fb}{{\rm fb}}
\newcommand{\pb}{{\rm pb}}


\input epsf



\section{Introduction\label{sec:Introduction}}

Determining how to identify and interpret signals of new physics within a
rapidly accumulating store of LHC data has become one of the primary
challenges for particle phenomenology of late.  Particular
attention has been focused on those signals which can be resolved within
the first few $\mathrm{fb}^{-1}$ of integrated luminosity.  These include
signals due to new strongly-interacting particles, which can be produced copiously 
at hadron colliders via strong interactions.  Since $SU(3)_c$ gauge invariance requires
each of these particles to decay down to a final state including one or more 
Standard-Model (SM) quarks or gluons, theories which predict such new particles 
frequently lead to excesses in multi-jet events --- excesses which can only be 
resolved above the sizeable SM QCD background via the application of astutely 
chosen event-selection criteria.

Nevertheless, while excesses in multi-jet events are a common prediction in
models of new physics, the optimal search strategy\footnote{The phrase 
``search strategy'' has a range of meanings in common use.  Here and throughout 
this paper, we use the phrase specifically to refer to the identification of 
particular channels relevant for the observation of new physics in the context 
of some extension of the Standard Model and the application of a particular set 
of event-selection criteria in order to resolve a signal of that new physics 
in those relevant channels.} for resolving such new physics above a sizeable 
SM background depends on the features of the model.  
One strategy useful in a number of beyond-the-Standard-Model (BSM) contexts 
is to focus on channels in which additional particles, such as
charged leptons, appear alongside the jets in the final state.   
Another is to search for multi-jet resonances arising from
the decay of new strongly-interacting particles.  Indeed, the utility of this
strategy has been demonstrated in a variety of BSM
contexts~\cite{SehkarMultijet,EssigThesis,HigherReps}, and has recently been
implemented, for example, in the trijet searches conducted by the
CDF~\cite{CDFTrijetSearch} and CMS~\cite{CMSTrijetSearch} collaborations.
Yet another fruitful strategy for resolving signals of new physics in
multi-jet events is to search for events with substantial missing transverse
energy ($\met$).  This strategy is particularly relevant in extensions of
the SM which include not only additional strongly-interacting fields, but
also a stable dark-matter candidate.
In traditional dark-matter models, the stability of the
dark matter candidate is frequently guaranteed by some symmetry, 
such as R-parity in supersymmetric theories, Kaluza-Klein parity~\cite{KKParity} in 
models with universal extra dimensions~\cite{Antoniadis,DDGLargeED,UED} (UED), or
T-parity~\cite{TParity} in little-Higgs theories~\cite{LittleHiggs}.
Any other, heavier particle charged under the same symmetry, once produced,
inevitably decays to a final state including one or more dark-matter particles,
which manifest themselves at colliders as $\met$.
By contrast, only a minute fraction of events produced by
pure QCD processes, which provide the dominant SM background for multi-jet events,
include substantial $\met$.  This quantity therefore
provides a useful discriminant between signal and background for multi-jet events
in a variety of new physics models which include a stable dark-matter candidate
(see, \eg, Ref.~\cite{TParity,ModelsWithJetsPlusMET}).  Searches for evidence 
of such models in the $\mathrm{jets} + \met$ channel have been performed
both at the Tevatron~\cite{D0SUSYSearch,CDFSUSYSearch,CDFDijetSearch} and at the
LHC~\cite{CMSSUSYSearch,ATLASSUSYSearchSimplifiedModel,AtlasJetsMET,CMSJetsMET}.
Moreover, even in alternative dark-matter scenarios
which do {\it not} include a single, stable dark-matter candidate~\cite{DDM}, 
and indeed even in certain extensions of the SM which do not relate directly 
to the dark matter problem, the $\met$ can still play a crucial role in 
resolving signals of new physics.

In this paper, we examine the prospects for resolving a particular class 
of models which give rise to signals in the $\mathrm{jets} + \met$ channel: 
those in which the additional strongly-interacting fields decay preferentially 
to SM states involving third-generation quarks, and in particular top quarks.  
Such ``top-rich'' scenarios arise in a number of BSM contexts and 
give rise to a variety of distinctive signature patterns.  Such signature
patterns play a crucial role in the LHC phenomenology of these scenarios --- 
especially in cases in which all relevant BSM fields decay essentially 
exclusively to states involving top quarks and invisible particles alone.   
A number of recent studies have assessed the prospects for detecting signals of
new physics in top-rich scenarios which give rise to the specific event topologies
$t\overline{t} + \met$~\cite{TwoTopsPlusMET,ShufangPlusThreeJets} and
$tt\overline{t}\overline{t} + \met$~\cite{TohariaGluino,MultiTopFromGluino,KaneTopChannel}.
Most of these studies have focused on search strategies which require the presence of 
one or more charged leptons in the final state, in addition to jets and $\met$. 
This approach has shown to be fruitful in many BSM contexts including 
supersymmetry (SUSY), in which the color-charged superpartners with complex decay 
chains generically produce final states of this sort.  

In this paper, we examine an alternative strategy for resolving signals
of new physics from the SM background in top-rich scenarios whose 
decay topologies are dominated by tops and invisible particles alone.     
This strategy involves focusing on the fully 
hadronic channel (\ie, on events which include no high-$p_T$ charged leptons), 
and selecting events primarily on the basis of two criteria: the number 
$N_j$ of high-$p_T$ jets in the event and the total missing transverse energy 
$\met$.\footnote{Top-rich scenarios also exist whose characteristic event topologies 
involve multiple top quarks, but no $\met$.  For such models, alternative search
strategies~\cite{FourTopsNoMET} are required.}
There are many advantages to this approach.  One of the principal ones 
is that the contribution to the SM background in the $\mathrm{jets} + \met$
channel from pure QCD processes arises due to ``fake'' sources 
of $\met$ (\eg, jet-energy mismeasurement), and several strategies 
(including those discussed in Refs.~\cite{Autofocus,JayJetMass}, as well as a
variety of data-driven techniques) exist for reducing 
this background to manageable levels.    
Moreover, the imposition of a charged-lepton veto can
significantly suppress the sizeable SM backgrounds from $t\overline{t} + \mathrm{jets}$
and $W+\mathrm{jets}$ events involving one or more leptonically-decaying $W$ bosons.
These events often contain substantial $\met$, due
to the presence of one or more
neutrinos in the final state, and consequently a substantial number of such events
survive stringent $\met$ cuts imposed to eliminate the QCD
background.  

In order to demonstrate the utility of this search strategy for
theories of this sort, we focus on a simplified model~\cite{SimplifiedModels} whose 
field content includes a color-octet fermion, a color-triplet scalar, and a 
color-singlet fermion which plays the role of the dark-matter particle.  This 
model can be realized within a certain limiting regime of
the minimal supersymmetric Standard Model (MSSM).  We find that
within the context of this example model, a search strategy based principally 
on $N_j$ and $\met$ turns out to be the optimal strategy for uncovering a signal 
of new physics at the LHC.  Our results  
can also be readily adapted to a wide variety of qualitatively similar models via an 
appropriate rescaling of the pair-production cross-sections and decays widths of the 
particles involved.  In many such cases, it is likely that a search based around these 
criteria will likewise be the optimal strategy for observing new physics.

Precedents for an analysis of this sort do exist in the literature in the 
context of particular models which give rise to top-rich event topologies, and it is 
encouraging to note that our claim is borne out in these cases.
In Ref.~\cite{ShufangPlusThreeJets}, for example, the discovery potential for new physics 
was assessed in a model containing an exotic quark $T'$ which decays directly to a 
top quark and a dark-matter particle with a branching fraction of effectively unity.  
It was shown that the best prospects for resolving a signal from the characteristic
$t\overline{t} + \met$ event topology which results from the pair-production of this $T'$ 
were indeed obtained in the fully hadronic channel when minimum cuts on $N_j$ and $\met$ 
were used as the primary event-selection criteria.  This result has been 
confirmed by a recent CDF analysis~\cite{CDFpsipsihadronic}.
The effectiveness of $N_j$ and $\met$ cuts in resolving signals of new physics has also been 
demonstrated in the context of no-scale $\mathcal{F}-SU(5)$
models~\cite{LiMaxin9jets1,LiMaxin9jets2}, and the utility of such cuts in conjunction 
with additional event selection criteria was also demonstrated in Ref.~\cite{Autofocus}.  
We observe that many of the analysis techniques pioneered in these works are in fact 
applicable to a broad class of BSM theories which predict signatures with large jet
multiplicities and substantial $\met$.

The outline of this paper is as follows.
In Sect.~\ref{sec:CrossSections}, we review the production and decay
properties of the gluino and lightest stop in our simplified model and discuss
the top-rich event topologies to which the model gives rise.
In Sect.~\ref{sec:CurrentConstraints}, we discuss the current limits on stop
and gluino masses from Tevatron and LHC data, taking care to distinguish between
those bounds which apply only when the masses of the first- and second-generation
squarks are light and those which apply even in the limit in which those masses
are taken to infinity.  In Sect.~\ref{sec:Backgrounds}, we enumerate the 
Standard Model backgrounds for a high-jet-multiplicity signal with substantial $\met$
and discuss the Monte-Carlo procedure adopted in our analysis.  
In Sect.~\ref{sec:Cuts}, we outline 
the program of event-selection criteria we employ in distinguishing such a signal
from those backgrounds.  In Sect.~\ref{sec:Results}, we present our results for
the LHC detection prospects for our simplified model and discuss how these results
may be extended to other, related scenarios which likewise give rise to top-rich
event topologies.  Finally, in Sect.~\ref{sec:Conclusions}, we conclude.

\section{Large Jet Multiplicities from Top-Rich Event
Topologies\label{sec:CrossSections}}

The primary purpose of this paper is to examine the prospects for observing 
signals of new physics in high-jet-multiplicity events at hadron colliders
afforded by adopting a search strategy focused on $N_j$ and $\met$.  As noted above,    
this search strategy is of particular interest in top-rich scenarios and other
theories whose decay phenomenologies are dominated by final states involving 
large numbers of high-$p_T$ jets and invisible particles.  In particular, our aim is to 
demonstrate the discovery reach which such a strategy is capable of achieving in 
scenarios of this sort within the first few $\mathrm{fb}^{-1}$ of integrated luminosity 
at the LHC, and to compare this reach to that afforded by alternative strategies.
We work primarily within the context of a simplified model whose 
BSM field content includes only three new particles: a color-octet Majorana fermion,
a color-triplet scalar, and a neutral, color-singlet fermion.  These particles
are assumed to transform under the symmetries of the theory identically to a 
gluino $\wtg$, a single, effectively right-handed stop $\st1\approx \widetilde{t}_R$, 
and a light, bino-like neutralino $\N1 \approx \widetilde{B}$ in the $R$-parity-conserving
MSSM.  For sake of convenience and to 
make contact with experimental limits extant in the literature, we shall henceforth 
refer to these particles by the corresponding sparticle names.  However, our results 
apply to any theory with the same field content and charge assignments, whether 
or not that theory arises in the context of SUSY.  Furthermore, we
assume a mass hierarchy $M_{\wtg}, \mst1 > m_{\N1}$ among these particles in what follows, 
and that $\N1$ is stable.  

One of the primary advantages of adopting such a toy theory as an
arena in which to study large-jet-multiplicity physics is that the decay behavior 
of both $\wtg$ and $\st1$ in this model is particularly simple.  Under the 
assumption of minimal flavor violation, $\st1$ decays almost exclusively via 
$\st1\rightarrow t \N1$ as long as this channel is kinematically open.  Similarly, 
$\wtg$ decays almost exclusively through 
$\wtg \rightarrow \st1 t \rightarrow t\overline{t}\N1$ and the conjugate process,
where $\st1$ may be real or virtual, depending on the relationship between
$M_{\wtg}$ and $\mst1$.  As a consequence of this simple decay phenomenology, 
each production process
which contributes to the signal-event rate in the $\mathrm{jets} + \met$ channel
is characterized by a {\it distinctive} distribution of jet multiplicities.
This example model therefore provides an excellent venue in which to assess the merits and
drawbacks of a jet-multiplicity-based event-selection strategy for a variety of 
qualitatively different production and decay scenarios by scanning over the 
model-parameter space.  Indeed, the qualitative results we obtain in the context of 
this model should be applicable in a wide variety of related scenarios characterized
by similar decay phenomenologies and can therefore serve as a useful rubric for 
assessing whether a jet-multiplicity-based search strategy is the optimal strategy 
for obtaining a signal of new physics in any particular such model.  

While our simplified model can be realized as a limiting case of the 
MSSM in which the masses of all other sparticles 
are infinitely heavy and decoupled, it is important to highlight the ways in which 
the collider phenomenology of this model differ from that realized in 
other regions of the parameter space of the general MSSM.  
With regard to the $\mathrm{jets} + \met$ channel, a crucial distinction 
arises between the regime in which the remaining squarks  
(hereafter collectively denoted $\wtq$) are heavy and decoupled and the regime in
which one or more of these squarks have TeV-scale masses.  Indeed,
as we shall see in Sect.~\ref{sec:CurrentConstraints}, experimental limits 
on sparticle masses (and on $M_{\wtg}$ in particular) differ significantly
between these two regimes.  It therefore behooves us to highlight the differences
between MSSM scenarios with and without light squarks from the perspective of
multi-jet collider phenomenology, before we move on to address $\wtg$ and 
$\st1$ production and decay in our toy theory.

The first significant distinction between SUSY scenarios with and without 
light $\wtg$ lies in the types of production processes which contribute
non-trivially to the total event rate in the $\mathrm{jets} + \met$ channel.
In models in which the $\wtq$ are decoupled, there are only two such processes
$pp\rightarrow \wtg\wtg$ and $pp\rightarrow \st1\st1^\ast$.  Indeed, the only other
contributions are those from $pp\rightarrow \wtg\st1$ and its conjugate process.
Under the assumption of minimal flavor violation, these contributions are suppressed
by small off-diagonal elements in the Cabibbo-Kobayashi-Maskawa (CKM) matrix and may
be safely neglected.  By contrast, in models with light $\wtq$, additional
pair-production processes such as $pp\rightarrow \wtg\wtq$ and
$pp\rightarrow \wtq\wtq^\ast$ also contribute.  As a result, the $N_j$ and
$\met$ distributions which result from such models can differ significantly from
those which result from models in which the $\wtq$ are heavy and decoupled.

\begin{figure}[ht!]
\begin{center}
  \epsfxsize 1.5 truein \epsfbox {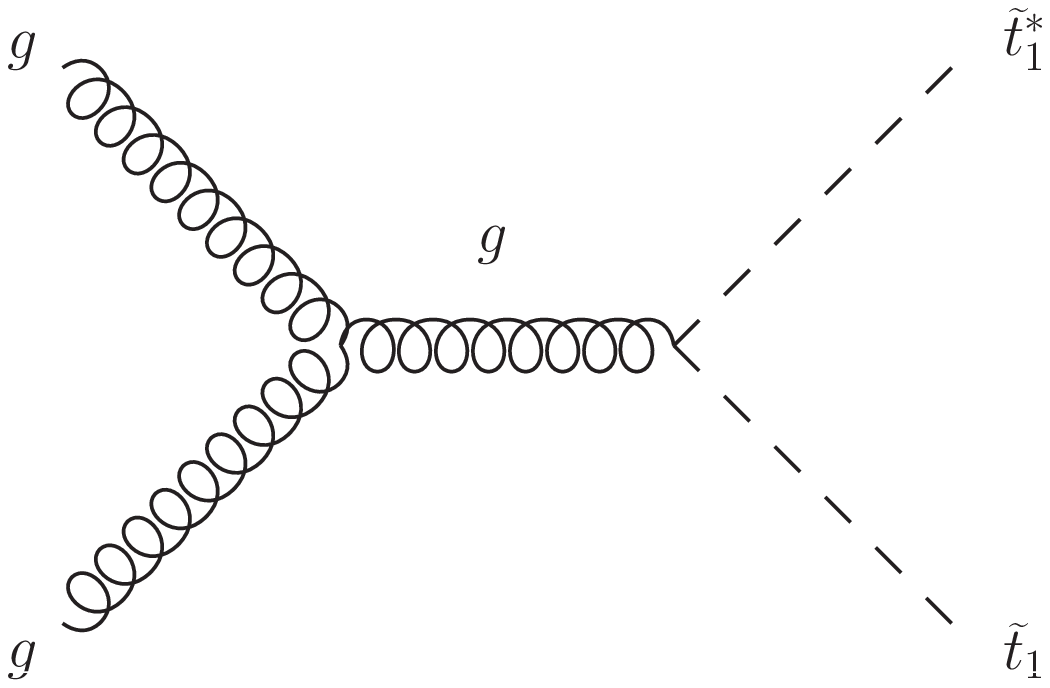}~~
  \epsfxsize 1.5 truein \epsfbox {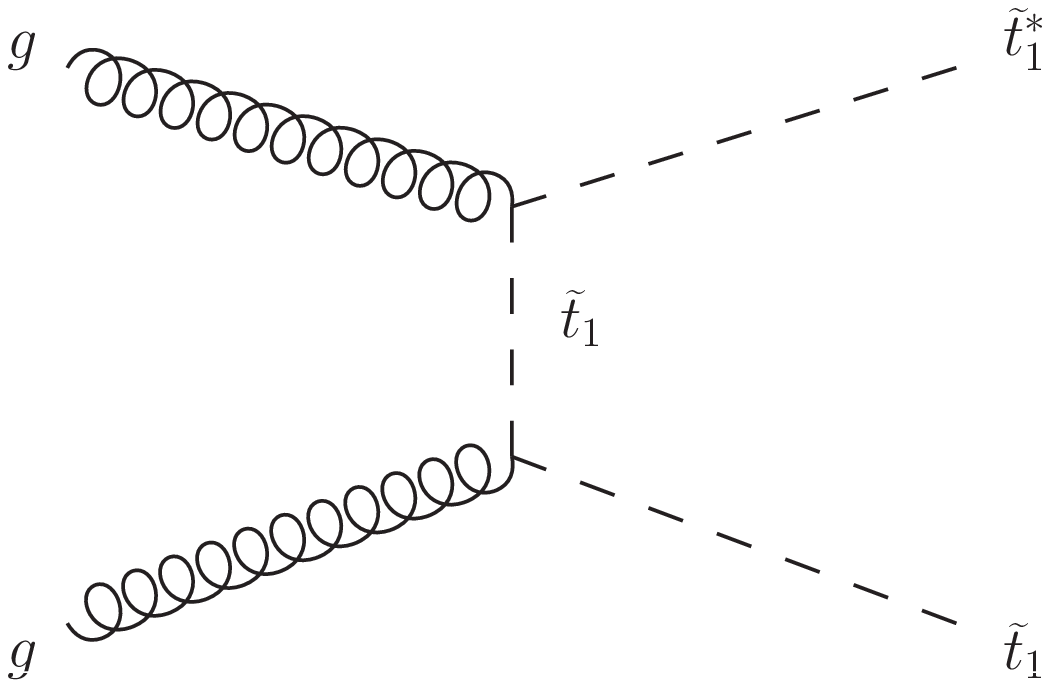}~~
  \epsfxsize 1.5 truein \epsfbox {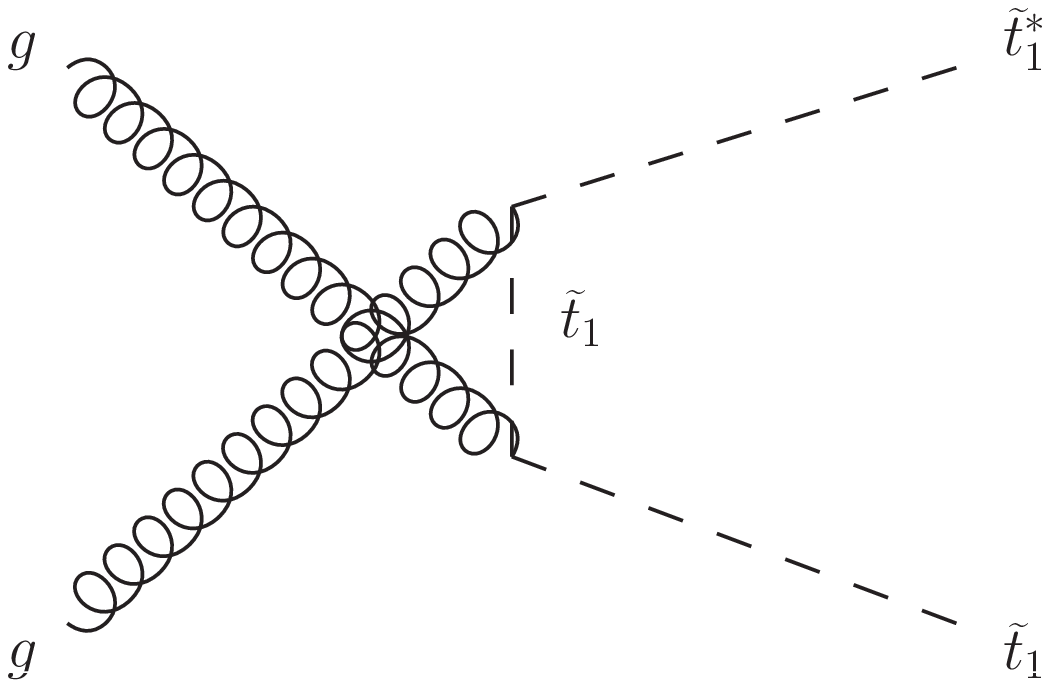}~~
  \epsfxsize 1.5 truein \epsfbox {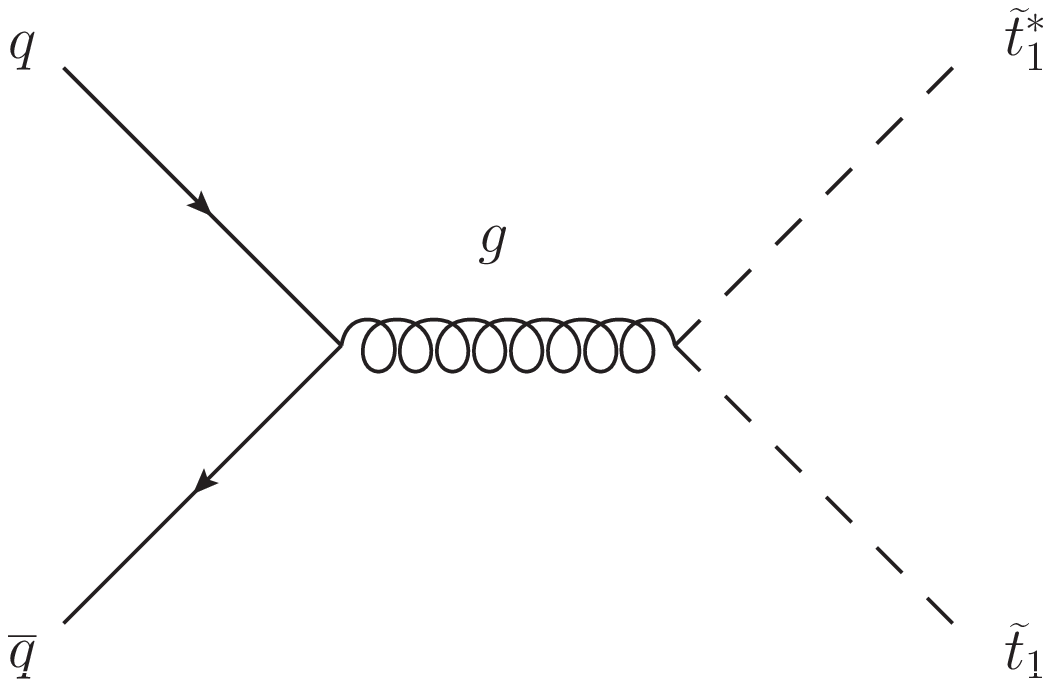}\\
  \vrule width 0.0pt height 0.5cm depth 0.5cm \\
  \epsfxsize 1.5 truein \epsfbox {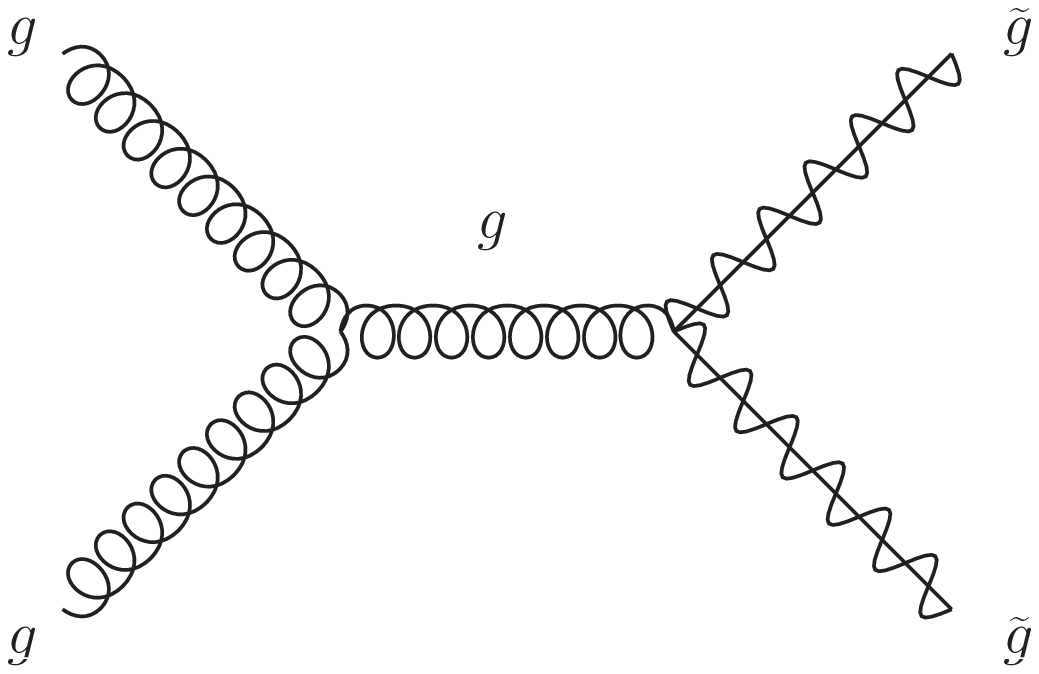}~~
  \epsfxsize 1.5 truein \epsfbox {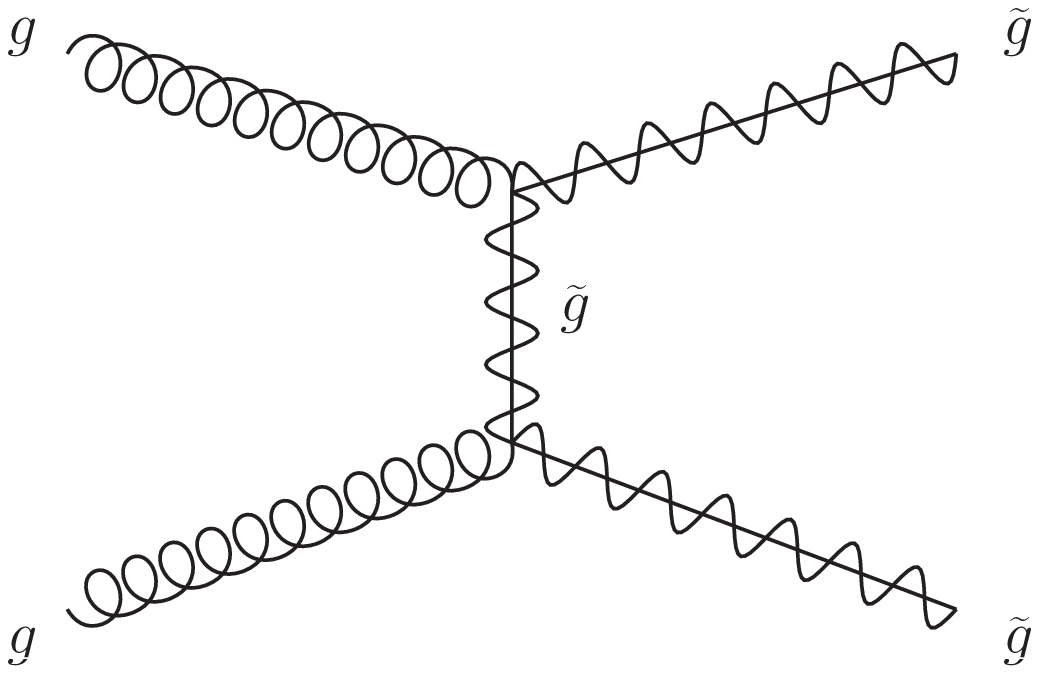}~~
  \epsfxsize 1.5 truein \epsfbox {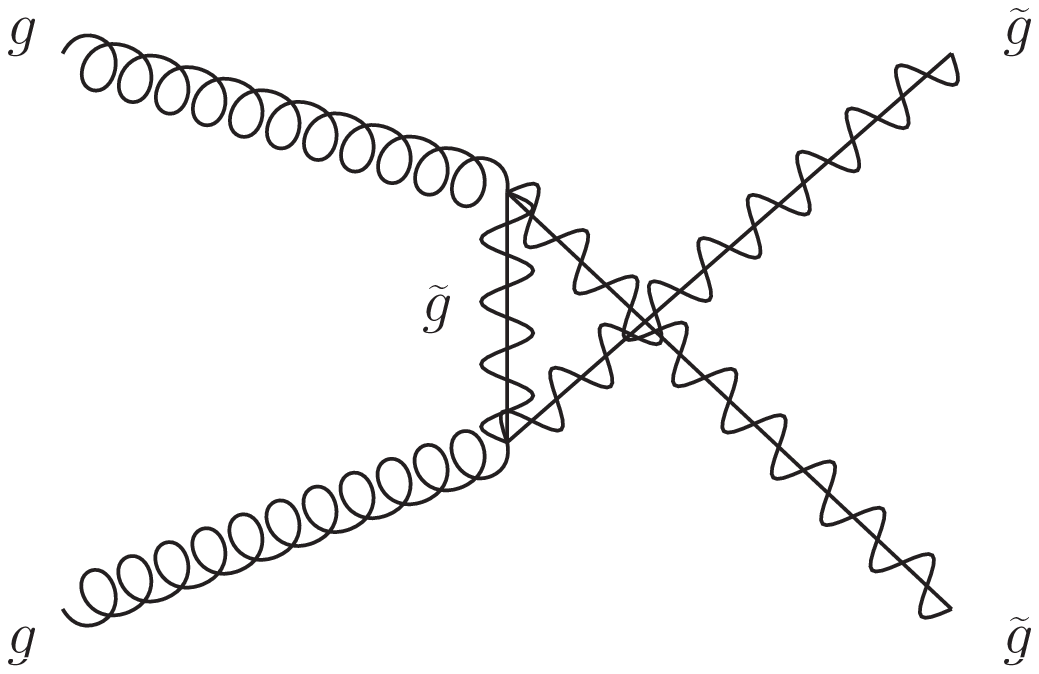}~~
  \epsfxsize 1.5 truein \epsfbox {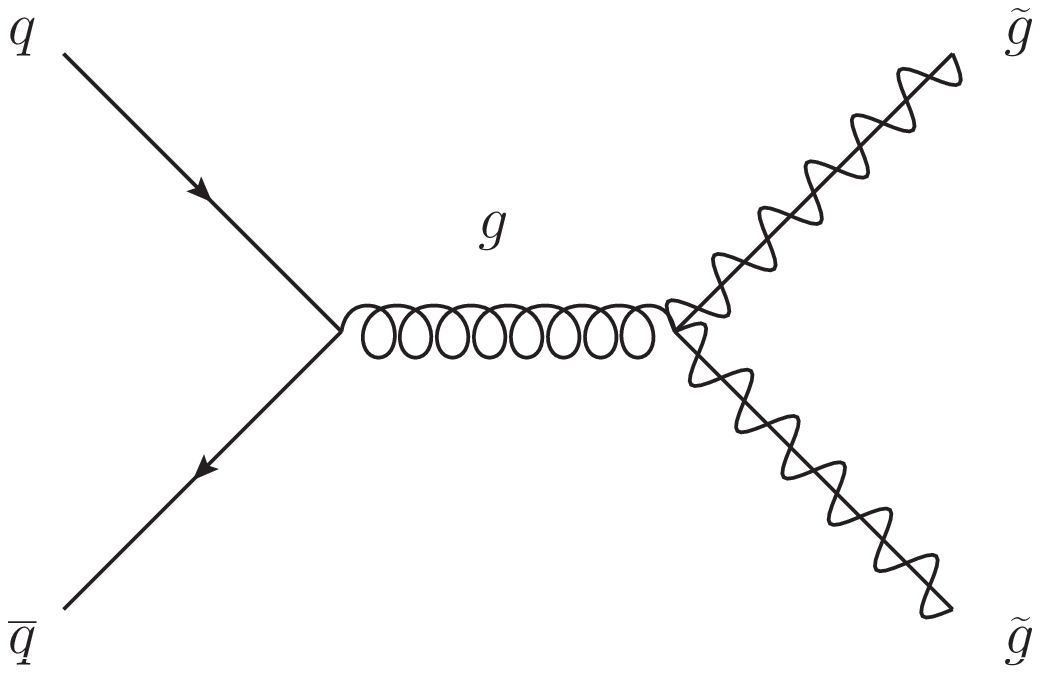}
\caption{Feynman diagrams associated with the processes relevant
for $\wtg$ and $\st1$ pair-production in the limit in which all
other squarks are taken to be heavy and decoupled.
\label{fig:FeynmanDiagrams}}
\end{center}
\end{figure}

A second difference between scenarios with and without light $\wtq$ arises 
due to the effect of these squarks on the cross-sections for those
multi-jet-production processes which occur in both sorts of models.
In decoupled models (again assuming minimal flavor violation), only the 
limited number of Feynman diagrams displayed in Fig.~\ref{fig:FeynmanDiagrams}
contribute to $pp\rightarrow \wtg\wtg$ and $pp\rightarrow \st1\st1^\ast$.  
By contrast, in models in which one or more of the first- and second-generation 
squarks are light, additional diagrams involving the $t$-channel exchange of 
such squarks also contribute, and the inclusion of these additional diagrams 
can result in a significant modification of the production cross-section for
$pp\rightarrow \wtg\wtg$ relative to that obtained in the decoupled-squark
limit.

A third significant difference between the collider phenomenologies associated with
decoupled and non-decoupled scenarios arises from differences in the
decay properties of the gluino in these two scenarios.
In the limit in which the $\wtq$ are decoupled, all decay chains initiated by
$\wtg$ decay necessarily involve either a real or virtual $\st1$, and hence lead to
the production of a $t\overline{t}$ pair.  Consequently, all fully hadronic final
states resulting from $pp\rightarrow \wtg\wtg$ production necessarily 
include at least twelve ``jets'' (\ie, quarks or gluons) at the parton level.  
This is generally not the case for models with light $\wtq$.  In particular, 
in scenarios in which $\mst1 > M_{\wtg}$, and no two-body decay channels are 
open for the gluino, the branching fractions for three-body gluino-decay processes 
involving virtual $\wtq$ (such as $\wtg\rightarrow q\overline{q} \N1$, where $q$ 
denotes any light quark) can be significant.  As a consequence, the expected jet 
multiplicities from $pp\rightarrow \wtg\wtg$ in the fully hadronic channel can often
be substantially lower in these scenarios than in those in which the $\wtq$ are
decoupled.  Moreover, we note that the decay phenomenology of both $\wtg$ and $\st1$ 
in the general MSSM are sensitive not only to the squark masses, but to the 
mass spectra of the chargino, neutralino, and slepton sectors as well.   

Having highlighted some of the salient differences between our example model
and other, similar models, we now examine its multi-jet phenomenology.
As discussed above, the only contributions to the total event rate in the
$\mathrm{jets} + \met$ channel in our toy theory arise due to gluino-pair and
stop-pair production.  The cross-sections for the partonic gluino-production processes
$gg\rightarrow \wtg\wtg$ and $q\overline{q}\rightarrow\wtg\wtg$ are
given at leading order (LO) by~\cite{DawsonXSec,KuleszaXSec}
\begin{eqnarray}
  \hat{\sigma}_{q\overline{q}\rightarrow \wtg \wtg}(\hat{s}) &=&
  \frac{8\pi\alpha_s^2}{9 \hat s}
  \left(1+  \frac{2M_{\wtg }^2}{\hat{s}}  \right) R \nonumber \\
  \hat{\sigma}_{gg\rightarrow \wtg \wtg }(\hat{s}) &=& \frac{3\pi\alpha_s^2}{4\hat{s}}\Bigg[
   3\bigg(1+\frac{4M_{\wtg}^2}{\hat{s}}-\frac{4M_{\wtg}^4}{\hat{s}^2}\bigg)
   \ln\left(\frac{1+R}{1-R}\right)-\bigg(4+\frac{17M_{\wtg}^2}{\hat{s}}\bigg)R
   \Bigg]~,
\label{eq:gogopartonicprocs}
\end{eqnarray}
where $ R \equiv \sqrt{1-4M_{\wtg}^2/\hat{s}}$.
Similarly, the partonic cross-sections for the corresponding stop-pair production
processes are given by~\cite{DawsonXSec}
\begin{eqnarray}
   \hat{\sigma}_{q\overline{q}\rightarrow \tilde t_1 \tilde t_1^*}(\hat{s}) &=&
      \frac{2\pi\alpha_s^2}{27\hat{s}}S^3 \nonumber \\
   \hat{\sigma}_{gg\rightarrow \tilde t_1 \tilde t_1^*}(\hat{s}) &=&
     \frac{\pi \alpha_s^2}{6\hat{s}^2}\left[
     \left(\frac{5}{8}+\frac{31m_{\tilde t_1}^2}{4\hat{s}}\right)\hat{s}S -
     \left(4 + \frac{m_{\tilde t_1}^2}{\hat{s}}\right)
     m_{\tilde t_1}^2\ln\left(\frac{1+S}{1-S}\right)
     \right]~,
    \label{eq:t1t1partonicprocs}
\end{eqnarray}
where $S \equiv \sqrt{1-4\mst1^2/\hat{s}}$.

\begin{figure}[ht!]
\begin{center}
\centerline{
  \epsfxsize 3.25 truein \epsfbox {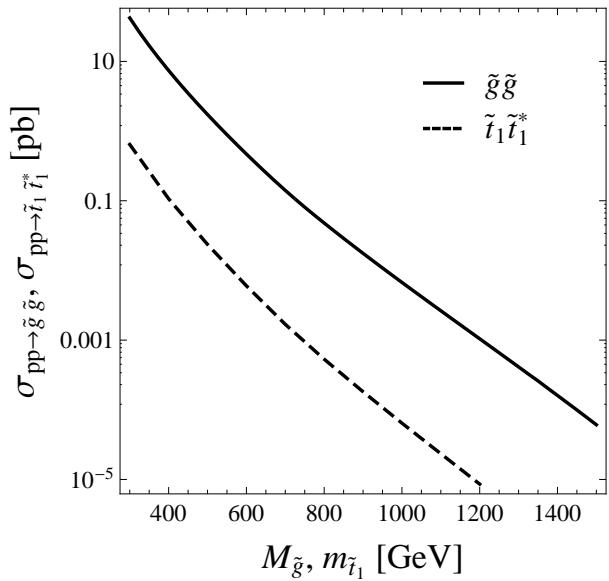} }
\caption{Total LO cross sections $\sigma_{pp\rightarrow\wtg\wtg}$ (solid curve) and
$\sigma_{pp\rightarrow\st1\st1^\ast}$ (dashed curve) at the $\sqrt{s}=7$~TeV LHC in
the limit in which all other squark masses are taken to infinity.
\label{fig:ProdXSecsPlot}}
\end{center}
\end{figure}

In Fig.~\ref{fig:ProdXSecsPlot}, we indicate how the total (LO) cross-sections
$\sigma_{pp\rightarrow \wtg \wtg}$ and $\sigma_{pp\rightarrow \st1 \st1^\ast}$
for gluino- and stop-pair production at the $\sqrt{s}=7$~TeV LHC behave as
functions of $M_{\wtg}$ and $\mst1$.  The curves appearing in this figure were
obtained by convolving the partonic cross-section formulae in
Eqs.~(\ref{eq:gogopartonicprocs}) and~(\ref{eq:t1t1partonicprocs}) with the
CTEQ6L1~\cite{CTEQ6PDFs} parton-distribution function (PDF) set and
approximating the running of $\alpha_s$ using the relation
\begin{eqnarray}
  \alpha_s(Q) &=& \alpha_s(M_Z)\Bigg(1+\frac{\alpha_s(M_Z)}{12\pi}
  \Bigg[23 - 2\Theta(Q^2-m_t^2)\ln\left(\frac{Q^2}{m_t^2}\right)
    -6\Theta(Q^2-M_{\wtg}^2)\ln\left(\frac{Q^2}{M_{\wtg}^2}\right)
    \nonumber\\ &&~~~~~~~~~~~~
    -\frac{1}{2}\Theta(Q^2-\mst1^2)\ln\left(\frac{Q^2}{m_{\st1}^2}\right)-
  \frac{11}{2}\Theta(Q^2-m_{\wtq}^2)\ln\left(\frac{Q^2}{m_{\wtq}^2}\right)\Bigg]
  \Bigg)^{-1}~,
  \label{eq:AlphaStrongRunning}
\end{eqnarray}
where $\Theta(x)$ is the Heaviside theta function, $Q$ is the energy scale at which the
coupling is being evaluated, and $m_t$ is the top-quark mass (taken here to be
$m_t=172$~GeV).  The fiducial value $\alpha_s(M_Z) = 0.130$ assigned to $\alpha_s$
at the weak scale was chosen to accord with the definition in the CTEQ6L1
PDF set.

\begin{figure}[ht!]
\centerline{
  \raisebox{0.1cm}{\epsfxsize 3.00 truein \epsfbox {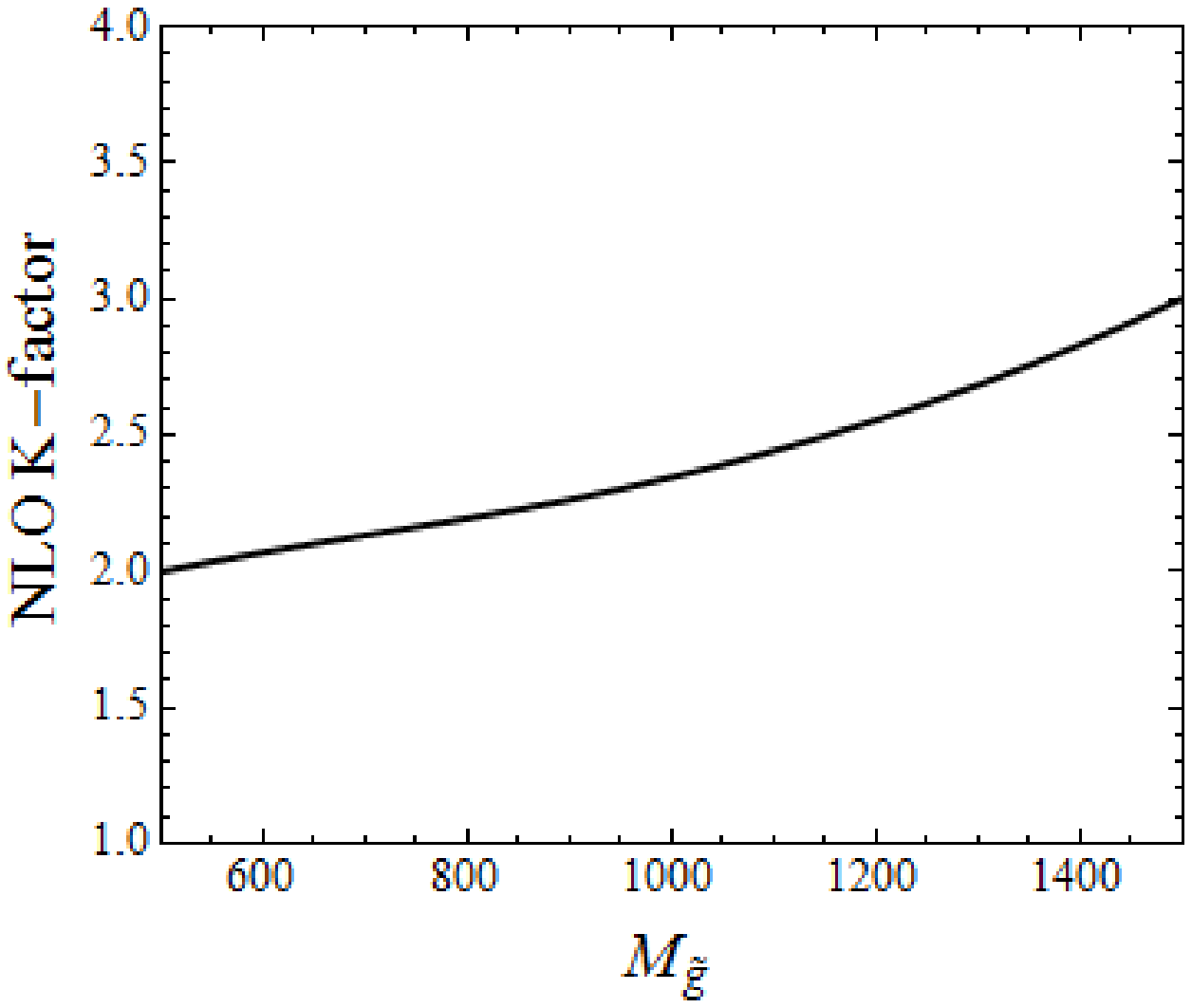}}
  \epsfxsize 3.25 truein \epsfbox {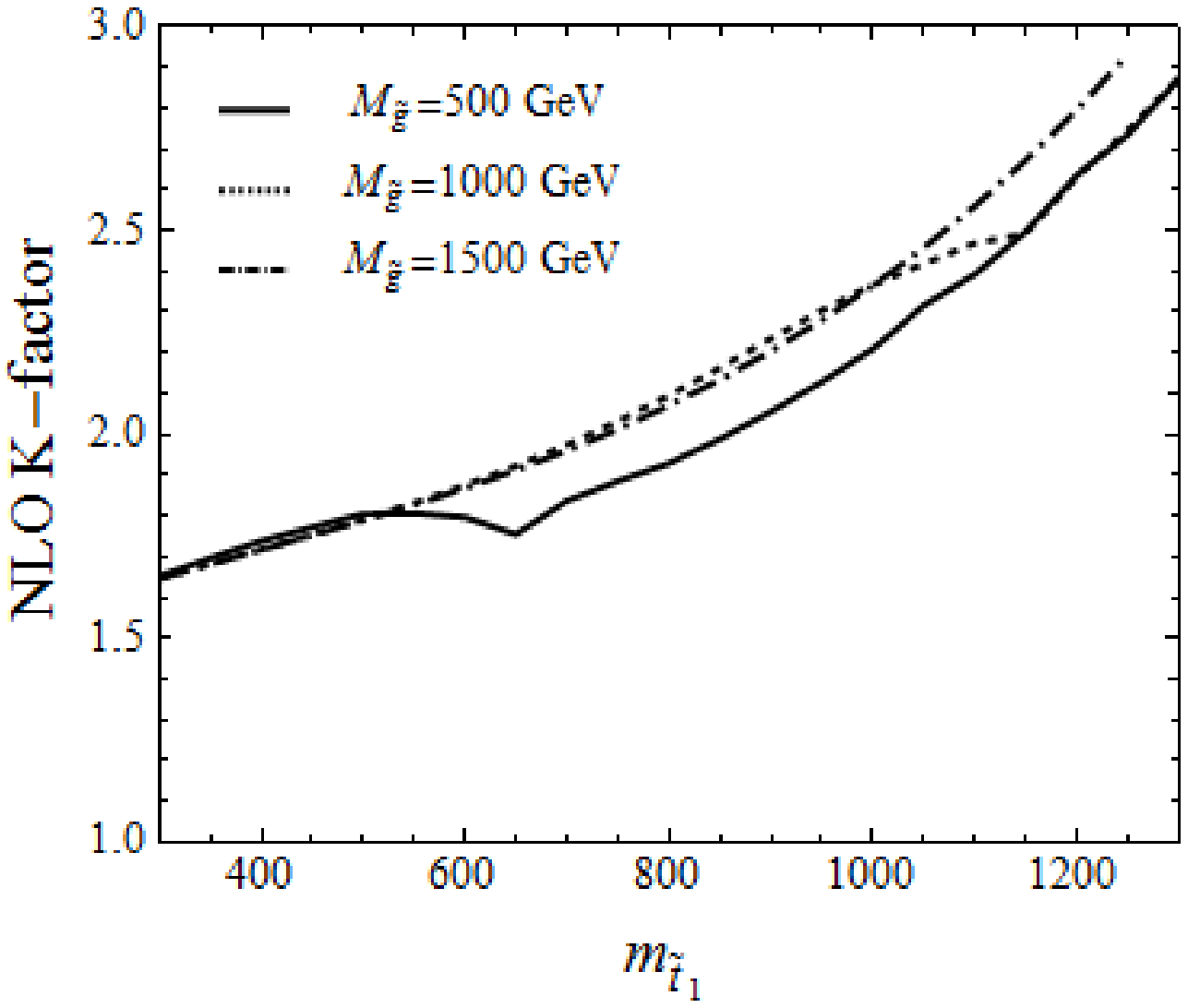} }
\caption{The NLO $K$-factors for gluino-pair production (left panel) and
stop-pair production (right panel) at a $\sqrt{s}=7$~TeV LHC,
shown as a function of the gluino mass $M_{\wtg}$ and stop mass $\mst1$,
respectively.  The results shown assume that all other squarks $\wtq$ are
infinitely heavy and decoupled.  The different curves in the right
panel correspond to different choices of $M_{\wtg}$.
\label{fig:NLOKFactor}}
\end{figure}

Since next-to-leading order (NLO) corrections to the
cross-sections for both $pp\rightarrow \wtg \wtg$ and $pp\rightarrow \st1 \st1^\ast$
production can be quite significant, we also account for the effects of
such corrections.  We approximate these effects by scaling our
leading-order results for $\sigma_{pp\rightarrow \wtg\wtg}$
and $\sigma_{pp\rightarrow\st1\st1^\ast}$ by the overall multiplicative factors
$K^{\mathrm{NLO}}_{\wtg\wtg}$ and $K^{\mathrm{NLO}}_{\st1\st1^\ast}$ 
shown in Fig.~\ref{fig:NLOKFactor},
which incorporate the effects of NLO modifications.  These NLO $K$-factors
were evaluated for each combination of $M_{\wtg}$ and $m_{\wtq}$ included in 
our analysis using the Prospino package~\cite{Prospino}.  
While the $K$-factor formalism employed here does not account for changes in
kinematical distributions of the decay products of the gluino, some information
on how these distributions are affected at NLO can be found in
Ref.~\cite{PlehnXSecCorrex}.  In this analysis, we do not include the effect of  
next-to-leading-log (NLL) corrections to these process from soft-gluon radiation, 
but we note that these corrections can provide an additional contribution as high
as $\mathcal{O}(20\%)$~\cite{BeenakkerNLONew} to the total production cross-section
in some cases.

\section{Current Constraints\label{sec:CurrentConstraints}}

In this section, we summarize the collider constraints on $M_{\wtg}$ and 
$\mst1$ in our example model.  We note that this summary not only 
delimits the viable parameter space of that model, but also 
provides a rough sense of the
discovery reach for top-rich scenarios afforded by the search strategies 
employed in the corresponding analyses.

A number of new-physics searches place limits on $M_{\wtg}$ and $\mst1$ in our
example scenario.  One example is the recent search
performed by the ATLAS collaboration~\cite{ATLASStopsMET}
at an integrated luminosity $\Lint = 35~\ipb$.  This analysis,
focused on events containing one lepton and two high-$p_T$ jets
(one of which was required to be $b$-tagged).  For $\mst1$ in the range
$130~\mathrm{GeV}\lesssim m_{\widetilde{t}_{1,2}}\lesssim 300$~GeV, the
collaboration obtained a limit $M_{\wtg} \geq 520~\gev$ on the gluino mass.
Likewise, limits on $\mst1$ can be derived from
the results of recent CDF searches for an exotic color-triplet fermion 
$\psi$ which decays preferentially to a top quark and a scalar dark-matter 
candidate $\chi$.  Search results in the semileptonic 
channel~\cite{CDFTprimeLimits} imply a bound $m_{\psi} \geq 360$~GeV on the 
mass of such a fermion, assuming $m_\chi \leq 100$~GeV, while corresponding 
search results in the fully hadronic channel~\cite{CDFpsipsihadronic} imply a 
slightly more stringent lower limit $m_\psi\gtrsim 380 - 400$~GeV for $m_\chi$ 
within this mass range (the precise lower bound depending on $m_\chi$).
By taking into account the difference between the pair-production cross-sections 
for scalars and fermions, one can translate this bound into a rough lower limit 
$\mst1 \gtrsim 285$~GeV in our example model, under the same assumption about 
$m_\chi$.  A similar search performed by the Atlas collaboration~\cite{AtlasttMET}
in the semileptonic channel at $\Lint = 35~\ipb$ yields
the constraints  $m_\psi \geq 275$~GeV and $m_\psi \geq 300$~GeV for
$m_\chi = 50$~GeV and $m_\chi = 10$~GeV, respectively.  Again accounting for 
the difference in production cross-section between scalars and fermions, 
we obtain the corresponding limits $\mst1 \gtrsim 190$~GeV and 
$\mst1 \gtrsim 215$~GeV in our example model for the 
corresponding choices of $m_\chi$.  A recent update~\cite{ATLASttMETRedux} 
with $\Lint = 1.04~\ifb$ has elevated this lower limit to
$\mst1 \gtrsim 270$~GeV for $m_\chi = 10$~GeV.
The CDF~\cite{Aaltonen:2010uf} and D\O\ \cite{Abazov:2008kz} collaborations 
have also performed searches for evidence of stop-pair production in the
fully leptonic channel at $\Lint = 1~\ifb$.  The non-observation
of a signal at either experiment implies a bound $\mst1 \geq 180$~GeV
(assuming that the dark matter particle is lighter than $100$~GeV).
CDF has conducted another search~\cite{Ivanov:2008st} at
$\Lint = 2.7~\ifb$
for $\widetilde{t} \widetilde{t}^\ast$ production followed by the
decay $\tilde t \rightarrow b \chi_1^\pm \rightarrow
b l \nu \chi_1^0$.  The results of this search imply a limit
$\mst1 \geq 150-185$~GeV in the context of our example model.
 
In addition to these constraints on top-rich scenarios, a number of other 
well-publicized limits on stop and gluino masses in SUSY theories 
(and in the MSSM in particular) had been derived from Tevatron and LHC data.  
For example, the ATLAS collaboration has searched
for evidence of squark and gluino production in final states involving
one isolated charged lepton, at least three high-$p_T$ jets
(with no $b$-tagging requirement), and $\met$~\cite{ATLASSquarksMET}.  
The results of this search, interpreted in the context of the constrained 
minimal supersymmetric Standard Model (CMSSM) with 
$A_0=0$, $\mu>0$ and $\tan \beta =3$, yield the constraint $M_{\wtg} \geq 700$~GeV.
A number of searches conducted at both the Tevatron and the LHC
in the $\mathrm{jets} + \met$ channel serve to constrain $M_{\wtg}$ as well.  
A recent CDF study~\cite{CDFSUSYSearch} of this sort, performed at 
$\Lint = 2~\ifb$, used the results of such a search to derive a limit 
$M_{\wtg} > 280~\gev$ in the context of minimal supergravity (mSUGRA)
in the $m_{\tilde q} \gg M_{\wtg}$ limit (in which gluino-pair 
production dominates the signal rate) for the mSUGRA parameter 
assignments $\mu <0$, $A_0 =0$, and $\tan \beta =5$.  
Corresponding D\O\ search results~\cite{D0SUSYSearch}, performed at a similar 
$\Lint$ and with similar mSUGRA parameter choices, imply   
a similar bound $M_{\wtg} > 308~\gev$.  An analogous study has also
been performed by the CMS collaboration~\cite{CMSSUSYSearch} at
$\Lint = 35~\mathrm{pb}^{-1}$.  Searches which do not assume a CMSSM or 
mSUGRA sparticle spectrum have also been performed at the LHC in the 
$\mathrm{jets} + \met$ channel.  A recent ATLAS study of this 
sort~\cite{AtlasJetsMET}, performed at $\Lint = 165~\ipb$, focused on the
case in which the third-generation squarks were taken 
to be extremely heavy, while the first- and second-generation squarks 
were taken to have TeV-scale masses.  For simplicity, the lightest 
neutralino was treated as massless.  For the case in which the first- and
second-generation squarks are reasonably heavy, with masses around $2$~TeV, 
a bound $M_{\wtg} \geq 725$~GeV was obtained.

One might assume that the results of these searches, which 
generally impose far more stringent limits than the constraints on top-rich
models quoted above, further serve to constrain the parameter space of our 
example model or place similar limits on other theories which also give 
rise almost exclusively to top-rich event topologies.  However,
these general SUSY results turn out not to be directly applicable to
our example model.  This is because
the sparticle-mass spectra and decay phenomenologies of other SUSY scenarios
often differ considerably from those obtained in that model, 
as discussed in Sect.~\ref{sec:CrossSections}.  For example, decay 
channels for squarks and gluinos which yield additional final-state 
leptons (beyond those produced from top decay) are often open in such 
scenarios.  Moreover, the $\wtq$ generally play a significant role in the collider 
phenomenology of general MSSM and CMSSM scenarios, whereas in our example model 
such squarks play no role whatsoever.  The stringent constraints on such scenarios 
we have quoted above are predicated on the effects of TeV-scale sparticles other than 
$\wtg$, $\st1$, and $\N1$.  In the absence of such effects, such constraints do 
not apply.  Likewise, the discovery reach associated with typical search 
strategies for squarks and gluinos is considerably lower.  However,
as we demonstrate below, a hadronic-channel analysis focused on $N_j$ and $\met$ 
can yield a significant increase in the discovery reach for top-rich 
scenarios whose collider phenomenology is dominated by final states involving
top quarks and $\met$ alone.

\section{Standard Model Backgrounds and Event Generation\label{sec:Backgrounds}}

We now discuss the SM processes which contribute to the background for 
large-jet-multiplicity searches in the $\mathrm{jets} + \met$ channel at the LHC.   
The dominant contributions come from processes such as $t\overline{t} + \mathrm{jets}$ 
and $W^\pm + \mathrm{jets}$, with all heavy particles decaying hadronically, and from
processes such as $Z + \mathrm{jets}$ and $t\overline{t}Z + \mathrm{jets}$
in which the $Z$ boson decays to a neutrino pair.  In addition, a sizeable
contribution arises from pure QCD processes in which the appearance of
substantial $\met$ arises due to jet-energy mismeasurement, and from 
$W \rightarrow \tau\nu_\tau$ decays where the $\tau$ is mistagged as a jet.  

In order to quantitatively assess the potential for resolving a signal of new physics above
these backgrounds in the context of our toy theory, we perform a Monte-Carlo analysis.  
In simulating the signal data, we fix $M_{\N1} = 100$~GeV and survey over a range of stop and
gluino masses from $300~\mathrm{GeV} \leq \mst1 \leq 1200$~GeV and
$500~\mathrm{GeV} \leq M_{\wtg} \leq 1500$~GeV in increments of $50$~GeV.
For each parameter-space point surveyed, a complete model was obtained using the
SuSpect 2.41 package~\cite{SuSpect}, and an NLO $K$-factor for each signal process
was obtained using Prospino~\cite{Prospino}, as
discussed in Sect.~\ref{sec:CrossSections}.
Events for both signal processes were then generated using the
MadGraph/MadEvent package~\cite{Madgraph} (with the CTEQ6L1~\cite{CTEQ6PDFs} PDF set)
for each point included in the survey and subsequently passed to
Pythia 6.4.22~\cite{PYTHIA} for fragmentation and
hadronization.  Realistic detector effects were simulated using
PGS4~\cite{pgs}, with detector parameters specified by the ATLAS detector card.

In this study, we do not simulate the QCD multi-jet background.  Instead, we
refer to the analyses performed in
Refs.~\cite{CutsKillQCDBGD0,CutsKillQCDBGATLAS},
which demonstrate that this background can be reduced to negligible levels
via the application of appropriate cuts on $N_j$, $\met$, and the
separation $\Delta \phi(p_{T_j},\displaystyle{\not} p_T)$ in azimuthal
angle between the most energetic jets and the missing-transverse-momentum
vector $\displaystyle{\not} p_T$.  In light of these results, we incorporate
a similar set of cuts into our preliminary event-selection criteria (\ie, the
``precuts'' detailed below) and assume that these cuts will likewise render
the QCD background negligible.  The remaining backgrounds from
$t\overline{t} + \mathrm{jets}$, $W^\pm + \mathrm{jets}$, $Z +\mathrm{jets}$,
and $t\overline{t}Z + \mathrm{jets}$ were explicitly simulated using the
Monte-Carlo procedure outlined above for signal-event generation, 
with matrix-element/parton-shower
matching applied for each process in the inclusive samples.
The $W^\pm + \mathrm{jets}$ and $Z +\mathrm{jets}$ background events were
generated assuming at least three jets present at the parton level.
While additional
processes such as $t\overline{t}t\overline{t} + \mathrm{jets}$ also contribute
to the total SM multi-jet background, the cross-sections for these
processes at the $\sqrt{s}=7$~TeV LHC are small enough that they may
be neglected.

\begin{table}
\begin{center}
\begin{tabular}{|c|c|ccc|}\hline
  ~Process~            & ~$\sigma_{\mathrm{LO}}$ (precuts)~
                       \vrule width 0.0pt height 0.5cm depth 0.3cm
                       & ~~$K_{\mathrm{NLO}}(\mu)$~~
                       & ~~~$\mu$~~~ &~~~$\sqrt{s}$~~~ \\ \hline
  ~$t\overline{t} + \mathrm{jets}$~&~ 267.2 fb ~& 1.40 & ~$m_t$~       & ~14 TeV~  \\
  ~$W^\pm + \mathrm{jets}$~        &~  60.4 fb ~& 0.65 & ~$M_W$~       & ~7 TeV~  \\
  ~$Z + \mathrm{jets}$~            &~  48.6 fb ~& 1.17 & ~$M_Z$~       & ~7 TeV~   \\
  ~$t\overline{t}Z +\mathrm{jets}$~&~   0.7 fb ~& 1.35 & ~$m_t+M_Z/2$~ & ~14 TeV~  \\ \hline
\end{tabular}
\caption{Leading-order cross-sections $\sigma_{\mathrm{LO}}$ for each of
the relevant SM background processes considered
in this analysis after the application of the precuts discussed in
Sect.~\ref{sec:Cuts}, as well as the NLO $K$-factor $K_{\mathrm{NLO}}(\mu)$ associated with
each process.  The factorization and renormalization scales at which each
$K$-factor has been evaluated are taken to be $\mu_F = \mu_R = \mu$, where the value
of $\mu$ for each process, along with the center-of-mass energy $\sqrt{s}$
for which each $K$-factor appearing in this table has been derived, are also
displayed.\label{tab:SMKFacs}}
\end{center}
\end{table}

Since NLO corrections to both $pp\rightarrow \wtg\wtg$ and
$pp\rightarrow\st1\st1^\ast$ can be significant,
such corrections must be applied to our background cross-sections as well.
As with the signal, we account for NLO corrections to the SM backgrounds using
the $K$-factor formalism.  Specifically, we scale the production cross-section
for each background process by the NLO $K$-factor associated with the partonic process
with the lowest jet multiplicity included in the Monte-Carlo simulation of that
background.  This procedure is expected to yield a conservative estimate of the 
overall signal significance, since $t\overline{t} + \mathrm{jets}$ provides the
dominant contribution to the SM background after a substantial cut on $N_j$ is
applied, and since the $K$-factors for the subprocesses which 
contribute toward the inclusive cross-section for 
$t\overline{t} + \mathrm{jets}$ tends to decrease as the number of jets 
involved increases.  The NLO $K$-factors for
several $Z + \mathrm{jets}$~\cite{KFactorsSMBergerZ} and
$W + \mathrm{jets}$~\cite{KFactorsSMBergerW} subprocesses
have recently been computed for the LHC at a center-of-mass energy
$\sqrt{s} = 7$~TeV.  While corresponding results for the remaining processes
included in our analysis are not yet extant in the literature for the same
center-of-mass energy,
$K$-factors for a number of $t\overline{t} + \mathrm{jets}$ subprocesses~\cite{KFactorsSMCampbell,KFactorsSMBern}, as well as for
$t\overline{t}Z + \mathrm{jets}$~\cite{KFactorttZ}, have been computed at
$\sqrt{s} = 14$~TeV.  We estimate the $K$-factors for these processes by 
adopting these results.
Numerical values for all $K$-factors used in this analysis are summarized in
Table~\ref{tab:SMKFacs}, along with the center-of-mass energy and common
factorization and renormalization scale $\mu$ at which each is evaluated.
The LO cross-sections for the corresponding background processes
(after the application of the precuts described in Sect.~\ref{sec:Cuts} below)
are also included in the table.

Estimates of the SM backgrounds obtained from Monte-Carlo 
simulations should always be taken with a grain of salt, especially when those 
estimates apply to regimes (such as the large-jet-multiplicity regime) 
for which little experimental data exist to corroborate them.
More accurate characterizations of the backgrounds relevant in these
regimes will come as more experimental data are accumulated.  However,
preliminary studies based on simulations of this sort are frequently 
invaluable, as it is often in such regimes that signals of new physics 
can be most readily resolved.  Moreover, the primary results of our analysis 
(for example, the projected discovery reach for $M_{\wtg}$ and $\mst1$) 
are not particularly sensitive to these uncertainties. 
  
\section{Surveying the Parameter Space\label{sec:Cuts}}

Having examined the individual signal and background processes which contribute
to the total event rate in the $\mathrm{jets} + \met$ channel in our example scenario,
we now outline the strategy we adopt for differentiating signal
from background events in this channel --- a strategy in which the principal
event-selection criteria are $N_j$ and $\met$.  In this analysis, $N_j$ is specifically 
defined to be the number of jets in the event with $p_T > 30$~GeV. 
As we shall see, this strategy actually renders the fully-hadronic 
channel even more auspicious for discovery in our example model 
than other, more conventional channels.  However, since our primary aim in this 
paper is to assess the efficacy of a search strategy based primarily 
on $N_j$ and $\met$ for top-rich scenarios in general, rather than this 
example model in particular,  we do not optimize our event-selection 
criteria for each combination of the model parameters $M_{\wtg}$, $\mst1$, and 
$M_{\N1}$; but rather introduce a small number of representative cutting 
regimens, each of which is particularly effective within a
certain characteristic region of model parameter space.
We emphasize that none of the cuts we impose as part of this program of
event-selection criteria involve $b$-tagging or top reconstruction.  Indeed, since
the dominant contribution to the SM background at large jet multiplicities is 
$t\overline{t} + \mathrm{jets}$, any further selection of
events based on these techniques would reduce signal and background
proportionally, and thus only serve to diminish signal significance.

As a first step in our event-selection procedure, we apply a preliminary set of
cuts, hereafter referred to as our ``precuts,'' to both the signal and background 
data.  These precuts are designed to mimic a realistic detector acceptance and 
to eliminate the sizable SM backgrounds from pure QCD processes, as discussed in Sect.~\ref{sec:Backgrounds}, and from low-jet-multiplicity events in general.  
In particular, we require that all events satisfy the following criteria:
\begin{itemize}
\item No isolated charged leptons ($e$ or $\mu$) in the final state.
\item At least five jets with $p_{T_j} > 40$~GeV.
\item Each jet must be isolated in the sense that $\Delta R_{jj} > 0.4$ for
      every possible pairing of jets in the event, where $\Delta R_{jj}$ is the
      separation distance in the $(\eta,\phi)$ plane between a given pair of jets.
\item An azimuthal-angle separation
      $\Delta \phi(p_{T_j},\displaystyle{\not}p_T) > 11.5^\circ$ between 
      the missing-transverse-momentum vector and each of the leading
      three jets (ranked by $p_{T_j}$).
\item $\met > 100$~GeV.
\end{itemize}

For the SM backgrounds, the only significant sources of missing energy are
neutrinos and jet-energy mismeasurement.  The first of these is suppressed quite
efficiently by the lepton veto, while the second is suppressed by the combination of
the minimum $\met$ cut and the requirement that the missing-transverse momentum vector
$\mptvec$ not be aligned (or anti-aligned) with any of the leading jets in a given
event --- the jets for which the potential for jet-energy mismeasurement is the
greatest.  It is worth noting that the $\Delta \phi(p_{T_j},\displaystyle{\not}p_T)$
cut employed here differs from the $\Delta\phi^\ast$ and $\alpha_T$ cuts employed
in Refs.~\cite{CMSAlphaT,CMSJetsMET} for a similar purpose.  As was pointed out
in Refs.~\cite{LiMaxin9jets1,LiMaxin9jets2}, these latter variables may not be
optimal for searches involving very large numbers of jets.  By contrast our $\Delta \phi(p_{T_j},\displaystyle{\not}p_T)$ cut,
excludes events based on the proximity of the $\displaystyle{\not}p_T$ vector to the
three highest-$p_T$ jets (for which the potential for jet-energy mismeasurement is the
greatest) rather than the nearest jet, and may therefore be a more reliable variable
in the large-jet-multiplicity regime.

\begin{figure}[ht!]
\begin{center}
\centerline{
  \epsfxsize 3.0 truein \epsfbox {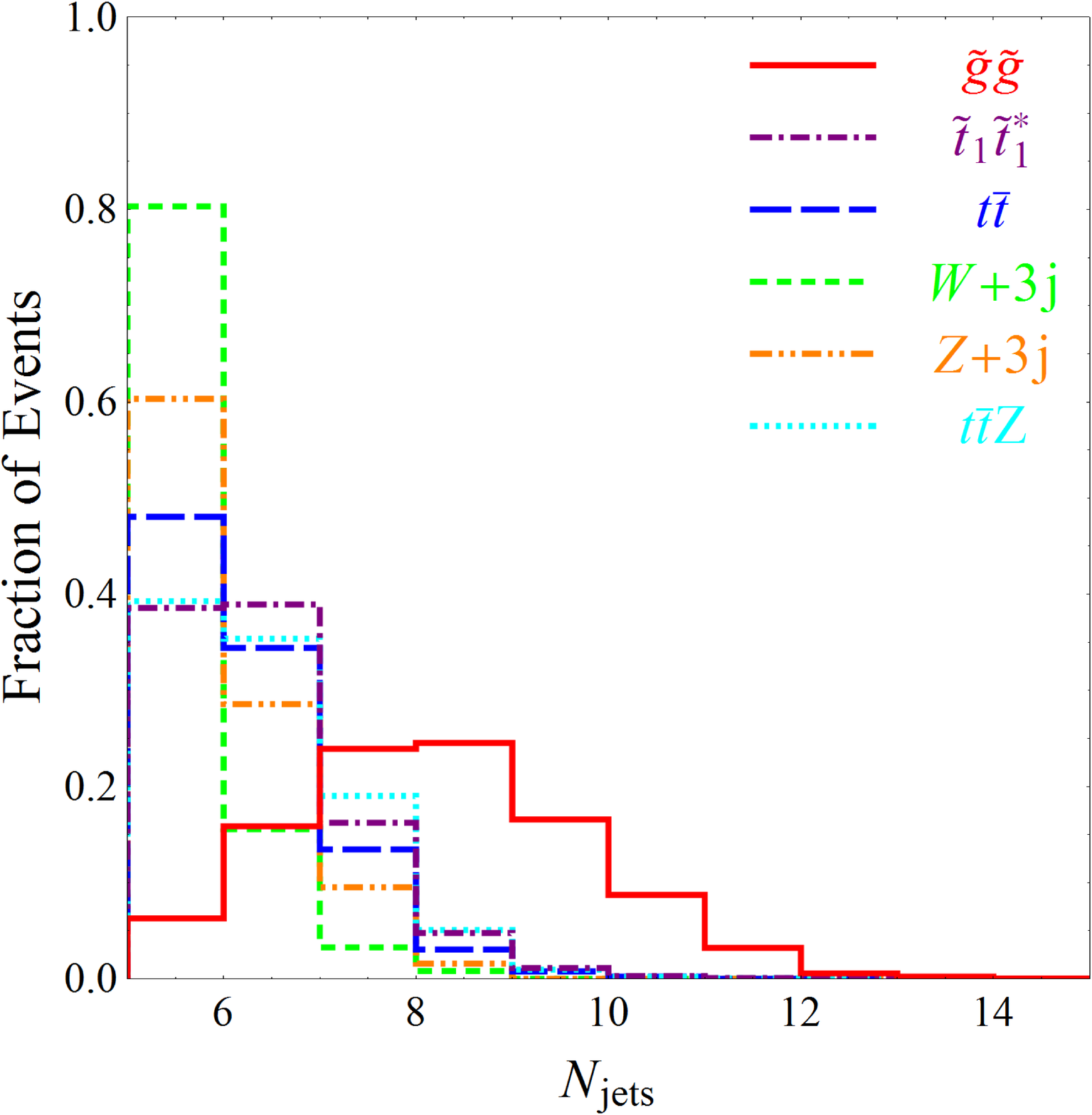} }
\caption{Event distributions for all relevant signal and background processes as
functions of jet number $N_j$ corresponding to the parameter choice
$M_{\wtg} = 1000$~GeV and $\mst1 = 600$~GeV in our simplified supersymmetric
model.  Each distribution shown has been normalized so that the total
area under it is unity.
\label{fig:JetDistribution}}
\end{center}
\end{figure}

\begin{figure}[ht!]
\begin{center}
\centerline{
  \epsfxsize 3.0 truein \epsfbox {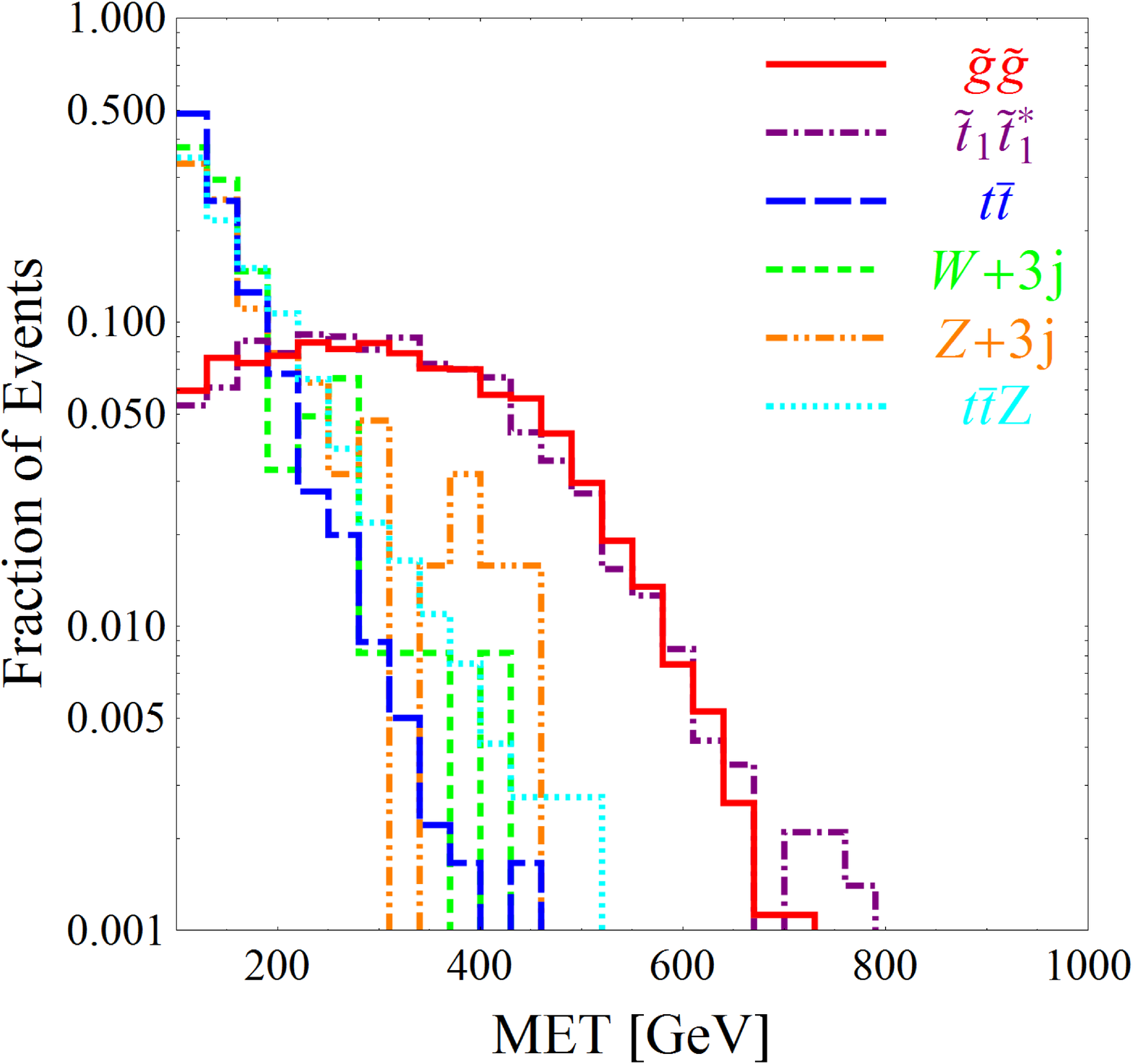}
  \epsfxsize 3.0 truein \epsfbox {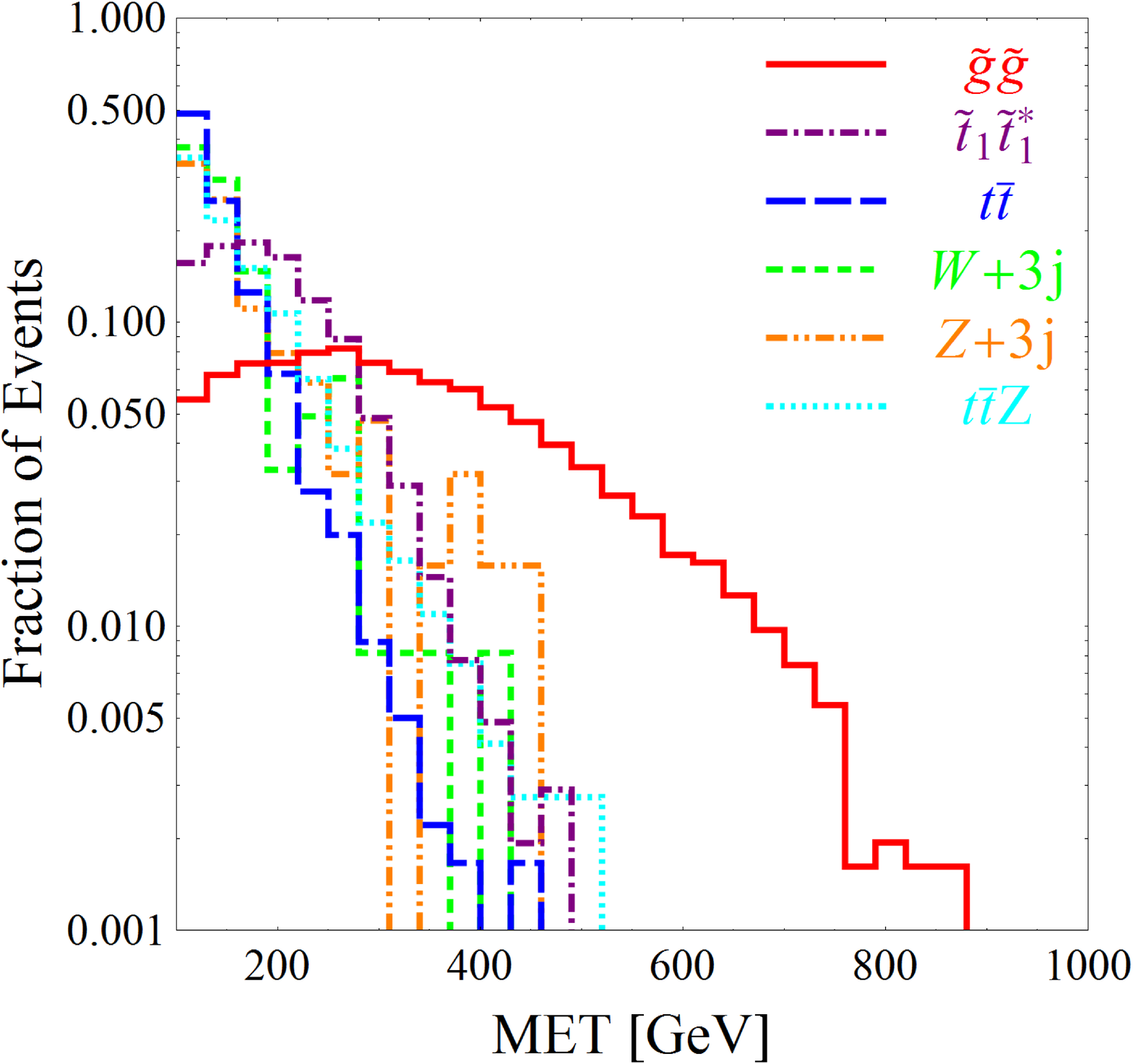} }
\caption{Event distributions for all relevant signal and background processes as
functions of missing transverse energy $\met$ for two illustrative combinations of
the gluino mass $M_{\wtg}$ and stop mass $\mst1$ in our simplified supersymmetric
model.  The distributions shown in the left panel correspond to the parameter choice
$M_{\wtg} = 800$~GeV and $\mst1 = 600$~GeV, while those shown in the right panel
correspond to $M_{\wtg} = 1200$~GeV and $\mst1 = 400$~GeV.
As in Fig.~\ref{fig:JetDistribution}, each distribution appearing in each panel
has been normalized such that the total area under each distribution is unity.
\label{fig:METDistribution}}
\end{center}
\end{figure}

We now consider the effect of imposing elevated cuts on $N_j$ and
$\met$ beyond those imposed in our precuts.  In order
to illustrate the effect that such additional cuts have on the statistical
significance for discovery in this example model, we display
the event distributions for all relevant signal and background processes as
functions of $N_j$ and $\met$ in Figs.~\ref{fig:JetDistribution}
and~\ref{fig:METDistribution} respectively.  In both of these figures, each
distribution shown has been normalized so that the total area under the
distribution is unity.

The results displayed in Fig.~\ref{fig:JetDistribution} correspond to
$M_{\wtg} = 1000$~GeV and $\mst1 = 600$~GeV; however, the shapes of the $N_j$
distributions for the signal processes do not vary significantly over the
range of parameter space surveyed, primarily because $N_j$ depends primarily
on the structure of the decay chain and is not particularly sensitive to the
details of the kinematics involved.  For example, as discussed in
Sect.~\ref{sec:CrossSections}, the characteristic number of ``jets'' at the
parton level (\ie, final-state quarks or gluons) expected in the
fully-hadronic channel from the $pp\rightarrow \wtg\wtg$ and
$pp\rightarrow \st1\st1^\ast$ signal processes is twelve and six,
respectively.  The characteristic jet multiplicity for
$pp\rightarrow \wtg\wtg$ at the detector level is reduced somewhat
because the sheer number of quarks and gluons in the final state makes
it increasingly likely that multiple such partons will be clustered together
into the same jet.  Nevertheless, this process clearly yields a substantial number
of events with large $N_j$.
We also remark that while the results shown in Fig.~\ref{fig:JetDistribution} 
correspond to a situation in which $M_{\wtg} > m_t + \mst1$, the $N_j$ 
distributions for $pp\rightarrow \wtg\wtg$ do not differ drastically from these
even in cases in which $M_{\wtg} < m_t + \mst1$ and gluino decay proceeds 
via an off-shell stop.
By contrast, the largest contributions to the SM background from
$t\overline{t} + \mathrm{jets}$ and $W^\pm + \mathrm{jets}$ after the application
of the precuts are those from events in which a $W^\pm$ boson decays to a $\tau$
lepton, which is misidentified as a jet, and a neutrino.
These backgrounds exhibit far lower characteristic jet multiplicities.  Indeed,
only a minute fraction of background events surviving our precuts
have $N_j \geq 8$, whereas roughly half of
the $pp\rightarrow \wtg\wtg$ events contain at least this number of jets.  This
figure therefore attests to how effective an elevated $N_j$ cut can be in
discriminating between signal and background in regions of parameter space
in which gluino-pair production provides the largest contribution to the signal.

The results shown in Fig.~\ref{fig:METDistribution} likewise motivate the
application of an elevated $\met$ cut in certain cases, although the utility
of such a cut is more sensitive to the values of $M_{\wtg}$ and $\mst1$.
The distributions displayed in the left panel of the figure correspond to
the parameter assignments $M_{\wtg} = 800$~GeV and $\mst1 = 600$~GeV.  For this
choice of parameters, as can be verified from Fig.~\ref{fig:ProdXSecsPlot},
gluino-pair production provides the dominant signal contribution, and the
$\met$ distributions for both $pp\rightarrow \wtg\wtg$ and 
$pp\rightarrow \st1\st1^\ast$ are sufficiently broad compared
to those for the SM background processes so as to render an elevated
$\met$ cut an effective discriminant between signal
and background.  By contrast, the situation displayed in the right panel,
which corresponds to the parameter assignments $M_{\wtg} = 1200$~GeV and
$\mst1 = 400$~GeV, is quite different.  Since $\st1$ is quite light in
this case, $pp\rightarrow \st1\st1^\ast$ provides the dominant signal
contribution.  However, another consequence of $\st1$ being so light is
that the $\N1$ produced by stop decay are not particularly energetic.
As a result, the $\met$ distribution for $pp\rightarrow \st1\st1^\ast$
does not differ as significantly from that of the SM background.
We therefore conclude that there is little to be gained by imposing
an elevated $\met$ cut in regions of parameter space in which $\mst1$
is small.  On the other hand, an elevated $\met$ cut does enhance detection
prospects for regions of parameter space in which the kinematics of $\st1$ or 
$\wtg$ decay is such that substantial kinetic energy is transfered to the $\N1$.

\begin{table}
\begin{center}
\begin{tabular}{|c|cc|cc|cccc|} \hline
\multirow{3}{*}{~~~Cuts~~~}
& \multicolumn{2}{|c|}{~$M_{\wtg}= 1000$~GeV~}
& \multicolumn{2}{|c|}{~$M_{\wtg}= 800$~GeV~}
& \multicolumn{4}{|c|}{~SM Backgrounds~} \\
& \multicolumn{2}{|c|}{~$\mst1 = 350$~GeV~} 
& \multicolumn{2}{|c|}{~$\mst1 = 800$~GeV~} 
& $t\overline{t}$ & $W^\pm$ & $Z$ & $t\overline{t}Z$ \\ 
& $\wtg\wtg$ & $\st1\st1^\ast$ 
& $\wtg\wtg$ & $\st1\st1^\ast$ 
& $(+\mathrm{jets})$  
& $(+\mathrm{jets})$
& $(+\mathrm{jets})$  
& $(+\mathrm{jets})$ \\ \hline
~precuts only~& ~ 2.04 & 37.21 & 15.77 & 0.16 & 374.09 & 66.28 & 56.86 & 0.95 \\
~$N_j\geq 6$~ & ~ 1.93 & 23.85 & 15.09 & 0.10 & 194.33 & 13.04 & 22.56 & 0.58 \\
~$N_j\geq 7$~ & ~ 1.60 &  9.47 & 12.92 & 0.04 &  65.54 & 2.72 & 6.32 & 0.24\\
~$N_j\geq 8$~ & ~ 1.11 &  2.72 &  8.93 & 0.01 &  15.19 & 0.54 & 0.90 & 0.06 \\
~$N_j\geq 9$~ & ~ 0.64 &  0.37 &  5.01 &   0  &   3.75 & 0 & 0 & 0.01 \\ \hline
~$\met\geq 200$~GeV~ &            ~ 1.35 & 8.06 & 11.15 & 0.14 & 41.47 & 10.32 & 15.34 & 0.23 \\
~$N_j\geq 6$, $\met\geq 200$~GeV~~&~ 1.28 & 4.73 & 10.66 & 0.08 & 21.54 & 1.09  &  3.61 & 0.14 \\
~$N_j\geq 7$, $\met\geq 200$~GeV~~&~ 1.06 & 1.73 &  9.07 & 0.03 &  7.27 &   0   &  0.90 & 0.06 \\
~$N_j\geq 8$, $\met\geq 200$~GeV~~&~ 0.73,& 0.47 &  6.20 & 0.01 &  1.68 &   0   &    0  & 0.01 \\
~$N_j\geq 9$, $\met\geq 200$~GeV~~&~ 0.41 & 0.05 &  3.38 &  0   &  0.42 &   0   &    0  &   0  \\
\hline
~$\met\geq 300$~GeV~ &            ~ 0.71 & 0.80 & 5.93 & 0.11 & 4.60 &  1.63 &  5.42 & 0.05 \\
~$N_j\geq 6$, $\met\geq 300$~GeV~~&~ 0.67 & 0.47 & 5.65 & 0.07 & 2.39 &   0   &  0.90 & 0.03 \\
~$N_j\geq 7$, $\met\geq 300$~GeV~~&~ 0.55,& 0.19 & 4.79 & 0.03 & 0.81 &   0   &    0  & 0.01 \\
~$N_j\geq 8$, $\met\geq 300$~GeV~~&~ 0.36 & 0.09 & 3.26 & 0.01 & 0.19 &   0   &    0  &   0  \\ 
~$N_j\geq 9$, $\met\geq 300$~GeV~~&~ 0.20 & 0.05 & 1.76 &   0  & 0.05 &   0   &    0  &   0  \\ 
\hline
\end{tabular}
\caption{Cross-sections (in femtobarns) for the signal processes 
$pp\rightarrow\wtg\wtg$ and $pp\rightarrow\st1\st1^\ast$ in a pair of 
illustrative benchmark scenarios with different values of the model parameters 
$M_{\wtg}$ and $\mst1$, as well as for the SM backgrounds from 
$t\overline{t} +\mathrm{jets}$, $W^\pm +\mathrm{jets}$, 
$Z +\mathrm{jets}$, and $t\overline{t}Z + \mathrm{jets}$ after the 
application of various cuts.  For further details, see text.  
\label{tab:AppendixXSecs}}
\end{center}
\end{table}

In order to provide a more quantitative demonstration of the effects of $N_j$ and $\met$
cuts on signal and background data, we provide a roster of cross-sections 
for the various signal and background processes considered in our analysis 
after the application of different combinations of such cuts in Table~\ref{tab:AppendixXSecs}.
All cross-sections quoted in this table include the relevant NLO $K$-factors.  In order to
demonstrate how the effect of these cuts depends on $M_{\wtg}$ and $\mst1$, we present 
signal cross-sections for a pair of benchmark scenarios representative of the two principal 
types of multi-jet phenomenology to which our example model gives rise.  
For the first of these scenarios, for which $M_{\wtg} = 1000$~GeV and $\mst1 = 350$~GeV,
the cross-section for the $pp\rightarrow \st1\st1^\ast$ production process is large 
enough that this process dominates the event rate in the $\mathrm{jets} + \met$ channel.
For this scenario, moderate cuts on $N_j$ and $\met$ similar to those imposed as part of the 
precuts offer the best prospects for discovery.  For the second benchmark scenario, 
for which $M_{\wtg} = \mst1 = 800$~GeV, $pp\rightarrow \wtg\wtg$ production dominates, and
the best prospects for discovery are obtained by imposing elevated $N_j$ and $\met$ cuts
on the order of $N_j \geq 8$ and $\met \geq 300$~GeV.

\section{Prospects for Discovery\label{sec:Results}}

The effect of imposing elevated cuts on $N_j$ and $\met$ on the statistical
significance of discovery in different regimes of model-parameter space can
be seen in Fig.~\ref{fig:SignificancePanels}.  In this figure,
we present a series of contour
plots indicating the regions of $(M_{\wtg},\mst1)$ parameter space
within which statistically significant evidence of new physics can be
obtained at the $\sqrt{s}=7$~TeV LHC after the application of several
different sets of elevated $N_j$ and $\met$ cuts.  All results shown in
these panels assume an integrated luminosity
$\mathcal{L}_{\mathrm{int}}=10~\mathrm{fb}^{-1}$.
Since the number of both signal and background events surviving the
cuts we impose is often quite small, we calculate the confidence 
level for all values of $M_{\wtg}$ and $\mst1$ included in our parameter-space
survey using Poisson statistics.  The significance contours shown in 
each panel of Fig.~\ref{fig:SignificancePanels} are those contours for which 
the equivalent confidence level for a Gaussian distribution would
correspond to a $3\sigma$ or $5\sigma$ discovery.
We also require that the signal is not event-count limited,
in the sense that the expected number of signal events at this luminosity which
survive all cuts exceeds five.  The gray striped region demarcated by the dashed
contour indicates the region of parameter space in which this event-count
criterion is not satisfied.

\begin{figure}[ht!]
\centerline{
  \epsfxsize 3.0 truein \epsfbox {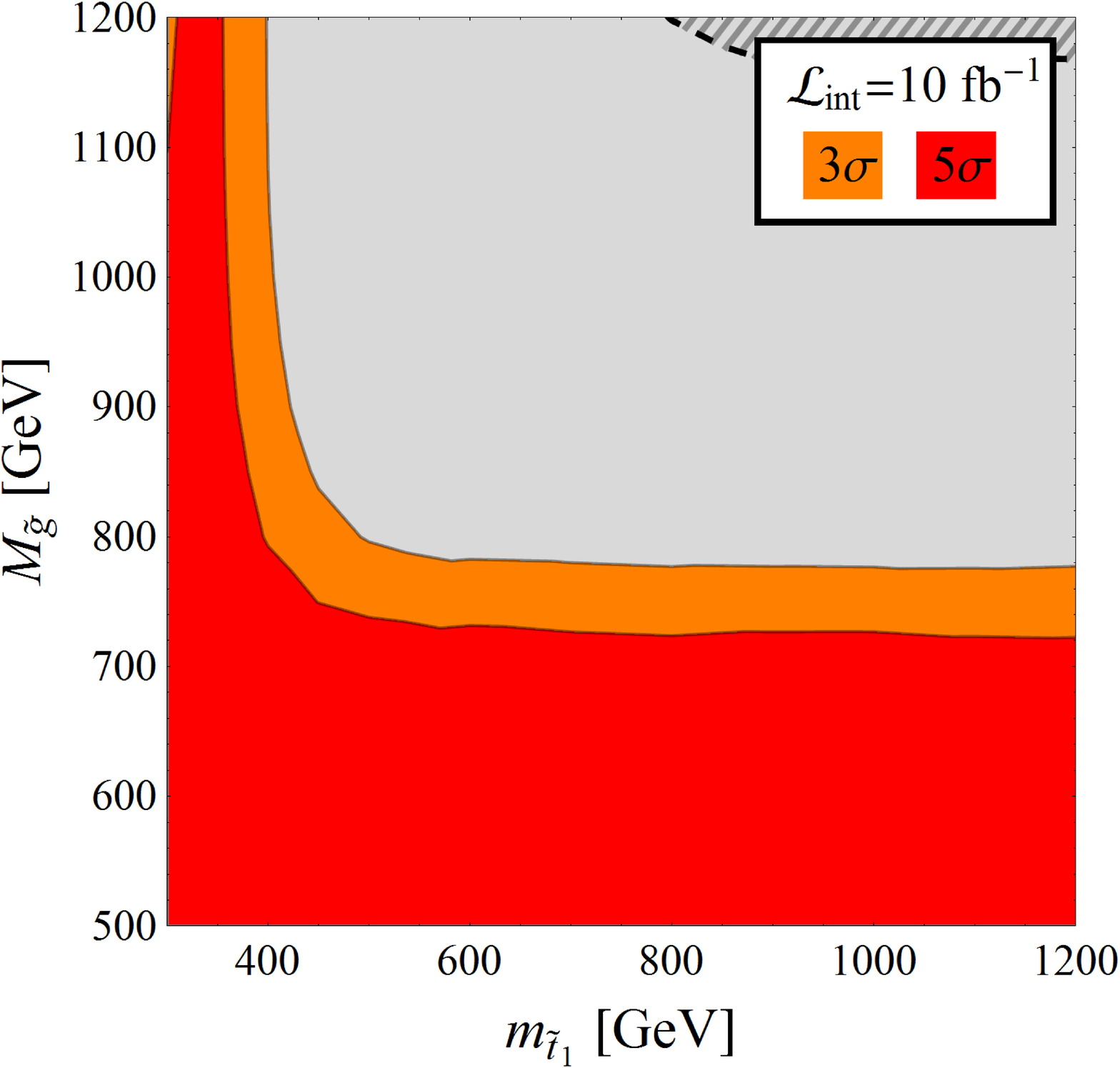}
  \epsfxsize 3.0 truein \epsfbox {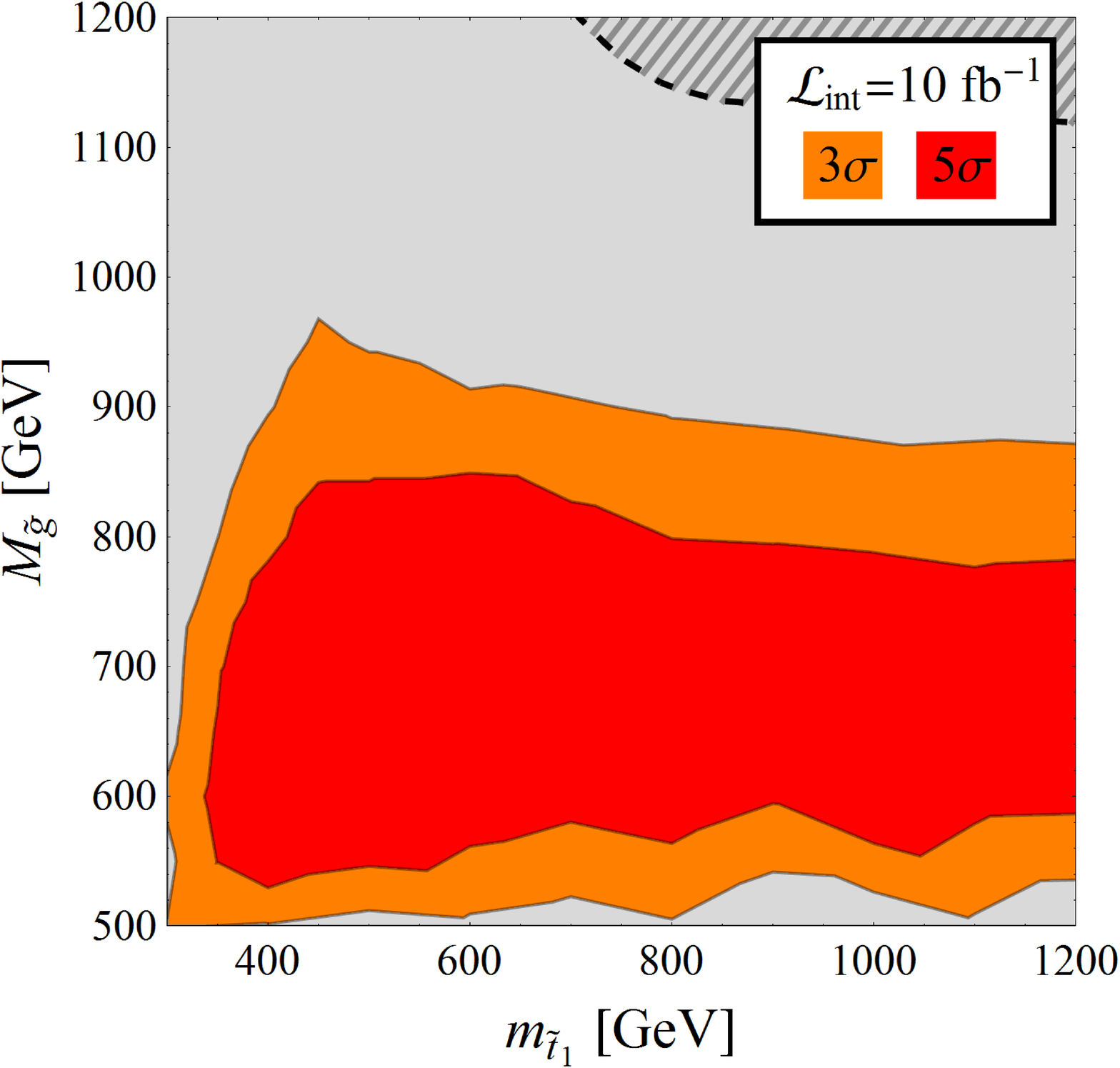} }
\centerline{
  \epsfxsize 3.0 truein \epsfbox {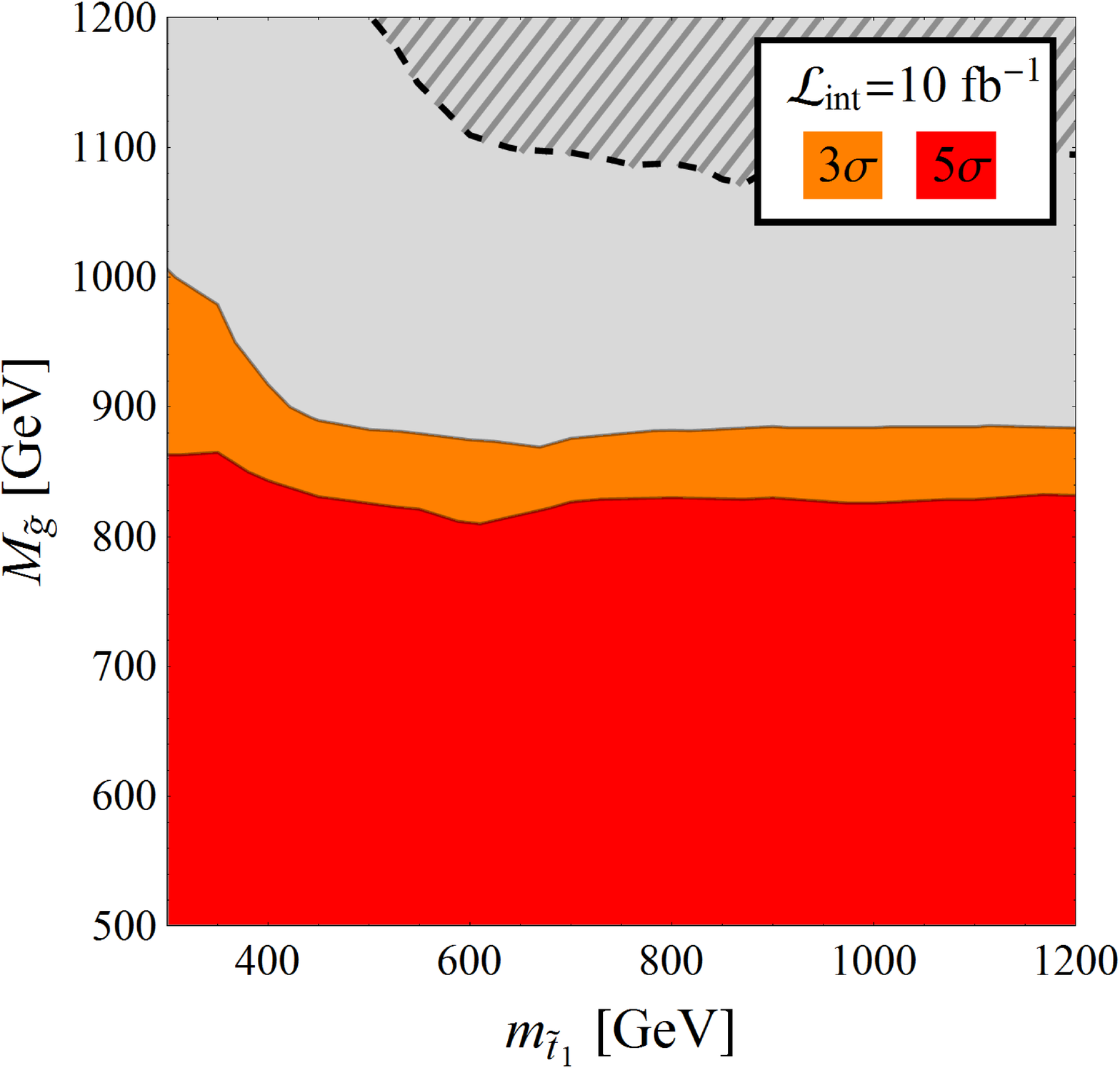}
  \epsfxsize 3.0 truein \epsfbox {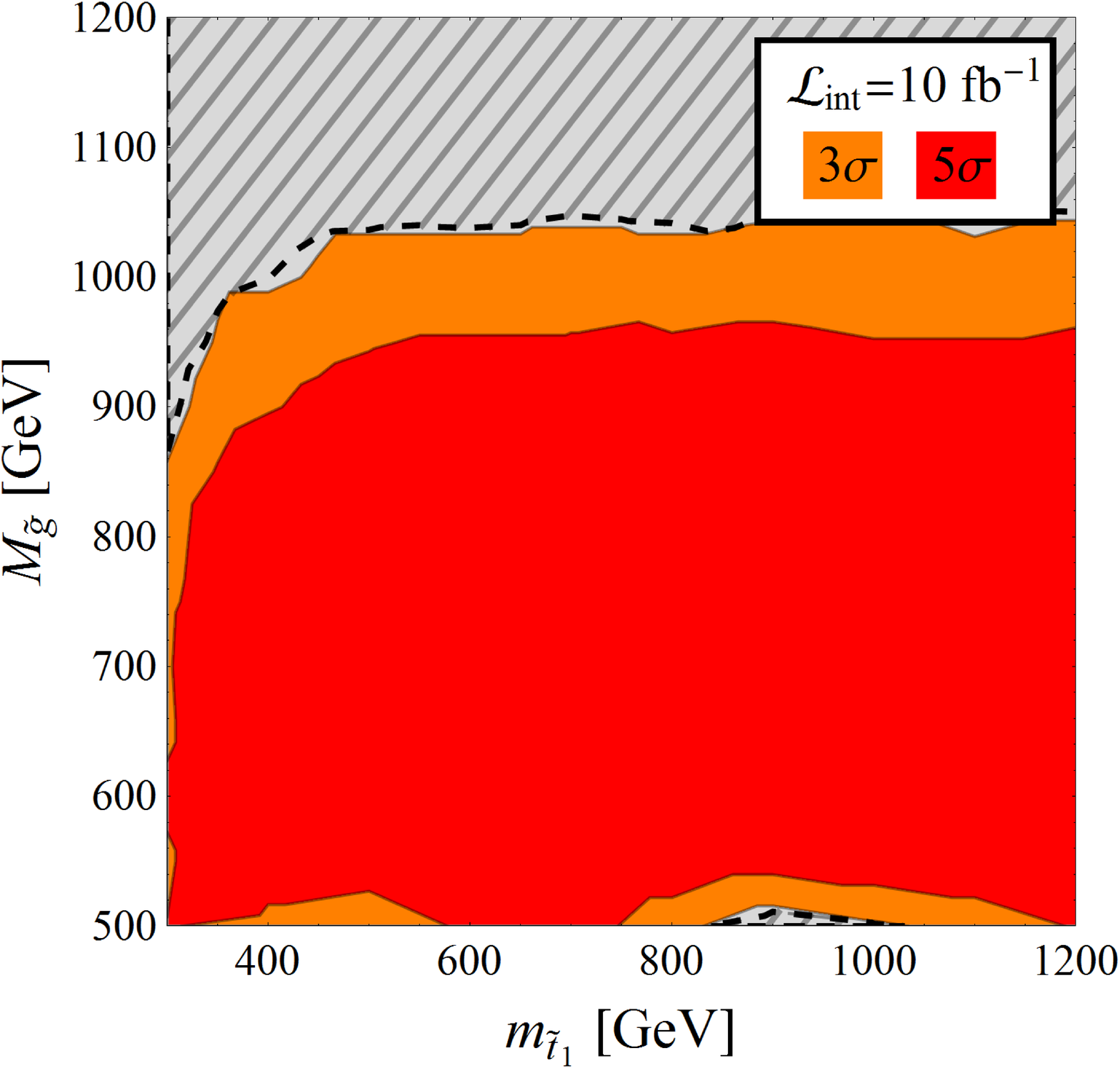} }
\caption{Contour plots illustrating the regions of $(M_{\wtg},\mst1)$ parameter space
within which evidence for new physics can be obtained at the $3\sigma$ (orange) or
$5\sigma$ (red) significance level with $\mathcal{L}_{\mathrm{int}} = 10~\mathrm{fb}^{-1}$
at the $\sqrt{s}=7$~TeV LHC after the application of the various sets of
cuts discussed in the text.  The gray striped region bounded by the dashed contour in each panel demarcates the
region within which the expected event count is less than five events at this luminosity.
The contours displayed in the upper left panel correspond to
the application of the precuts alone.  Those in the upper right are obtained by
imposing an additional $\met\geq 300$~GeV cut on top of those precuts, those in the lower
left panel are obtained by imposing an additional $N_j\geq 8$ cut on top of the
precuts, and those in the lower right panel are obtained by applying these two 
additional cuts in tandem.
\label{fig:SignificancePanels}}
\end{figure}

The contours displayed in the upper left panel of
Fig.~\ref{fig:SignificancePanels} are obtained by imposing
the precuts alone.  In this case, we see that
in the region of parameter space in which $\mst1 \lesssim 400$~GeV,
the signal contribution from $pp\rightarrow \st1\st1^\ast$ is
substantial, and the $N_j\geq 5$ and $p_{T_j}$ cuts imposed as part
of the precuts are sufficient to resolve the signal from this
process from the SM background.  Likewise, for $M_{\wtg} \lesssim 725$~GeV,
these same cuts alone are also sufficient to resolve the signal from the
$pp\rightarrow \wtg\wtg$ process.

The situation changes, however, when the minimum $\met$ threshold is elevated, 
as is evident from the upper right panel of Fig.~\ref{fig:SignificancePanels}.
This panel displays the results obtained by imposing an additional 
$\met \geq 300$~GeV cut on top of the $\met \geq 100$~GeV requirement included
as part of the precuts.  These results indicate that
in regions of parameter space in which $M_{\wtg} \gg 2 m_t + M_{\N1}$,
for which neutralinos produced via the $pp\rightarrow \wtg\wtg$ process
tend to be quite energetic, $\met$ serves as an extremely 
effective criterion for distinguishing signal from background, as we also saw in
Fig.~\ref{fig:METDistribution}.  (An exception occurs in cases in which
$M_{\wtg} \gg \mst1$ and the top quarks receive a greater proportion of
the mass energy of the decaying gluino.)
The same is true of neutralinos produced
via the $pp\rightarrow \st1\st1^\ast$ process in regions
of parameter space in which $\mst1 \gg m_t + M_{\N1}$.
Conversely, when
these conditions on $M_{\wtg}$ and $\mst1$ are not satisfied, $\met$ serves
as a poor discriminant between the respective $pp\rightarrow \wtg\wtg$ and
$pp\rightarrow \st1\st1^\ast$ signal contributions and the SM background, 
and the significance of discovery afforded by the corresponding signal 
process decreases.

The lower left panel of Fig.~\ref{fig:SignificancePanels} displays the
results obtained by imposing an additional $N_j \geq 8$ cut in addition 
to the precuts.
Events produced by $pp\rightarrow \wtg\wtg$, for which the characteristic
jet multiplicity is quite high, tend to survive such a cut, which is quite
efficient in eliminating events from the SM backgrounds.  For this reason,
the discovery potential afforded by this signal process is improved with 
each incremental increase in the $N_j$ requirement up to a reasonably high
threshold --- unless other considerations,
such as the total expected number of signal events, become an issue.
In accord with the results displayed in Fig.~\ref{fig:JetDistribution}, we 
find that this threshold occurs for a cut in the $N_j \gtrsim 8 - 10$ range.
By contrast, events produced by $pp\rightarrow \st1\st1^\ast$ tend to
involve far lower jet multiplicities; hence substantially elevating the $N_j$ cut
results in a loss of significance in those regions of parameter space within
which this process is responsible for the dominant signal contribution.

The lower right panel of Fig.~\ref{fig:SignificancePanels} displays the 
results obtained by imposing both an $N_j \geq 8$ cut and a $\met\geq 300$~GeV 
cut in addition to the precuts.  This set of cuts yields a $5\sigma$ 
significance throughout a broad region of parameter space within which
$550~\mathrm{GeV}\lesssim M_{\wtg}\lesssim 950~\mathrm{GeV}$.  This 
clearly constitutes a far greater reach than that obtained by imposing 
either an elevated $N_j$ cut or an elevated $\met$ cut alone.  As before,
the elevated $N_j$ cut results in the signal rate being dominated by 
gluino-pair production, as expected.  It is also evident from the results
displayed in this panel that at $\mathcal{L}_{\mathrm{int}}=10~\mathrm{fb}^{-1}$, 
the effect of the combined $\met$ and $N_j$ cuts imposed here on the 
signal-event rate is quite severe.  Indeed, we find that further increasing  
the jet-multiplicity threshold beyond $N_j \geq 8$ or substantially elevating 
the $\met$ cut above $\met \geq 300$~GeV does not result in a significant 
improvement in the reach, essentially because the signal becomes event-count
limited. 

In addition to the statistical significance of discovery itself, the 
signal-to-background ratio $S/B$ is also of interest, since 
estimates of that significance can be unreliable for large  
$S/B$ due to systematic uncertainties in the expected backgrounds.   
For each combination of elevated $N_j$ and $\met$ cuts  
considered in Fig.~\ref{fig:SignificancePanels},
we have verified that for $S/B \geq 0.1 $ throughout all 
regions of parameter space shown in that figure within
which a significance of $3\sigma$ or above is obtained.  This suggests 
that our results are indeed robust against systematic uncertainties affecting
the background.  Indeed, only for the precuts alone do we find that this
criterion on $S/B$ is not satisfied, and only within certain regions of 
parameter space within which $\mst1 \ll M_{\wtg}$.

One can compare these results to those afforded by other, complementary
strategies~\cite{TohariaGluino,KaneTopChannel} for observing a signal of
new physics from a pair of gluinos decaying into
$t\overline{t}t\overline{t} + \met$.  From among those strategies,
the best reach is obtained by demanding one lepton and four $b$-tagged jets in
the final state~\cite{KaneTopChannel}, which permits a $5\sigma$ discovery
for $M_{\wtg}$ as high as $650$~GeV at an integrated luminosity of
$1~\mathrm{fb}^{-1}$.  By contrast, with the search strategy we adopt here
--- and specifically by requiring that $N_j\geq 8$ and $\met \geq 200$~GeV --- we 
find that the LHC reach (at the same significance level and with the same
luminosity) can be increased to roughly $730$~GeV.  This serves as just one
example of how a search strategy focused on fully hadronic events with 
large jet multiplicities and substantial $\met$ can be useful in identifying
signals of new physics in top-rich scenarios at the LHC.  

To summarize the qualitative results of this section, we find that in situations 
in which $pp\rightarrow \wtg\wtg$ production dominates, 
focusing on high-jet-multiplicity events in the fully hadronic channel is an 
effective strategy for uncovering new physics at the LHC.  By contrast, we find 
that in situations in which $pp\rightarrow \st1\st1^\ast$ production (a process 
for which the characteristic jet multiplicities are far lower) dominates, this 
strategy is far less effective.  These results therefore provide insight into
what sorts of scenarios are likely to benefit from such a search strategy, and 
what sorts of models may be less amenable to such techniques.

\section{Conclusions\label{sec:Conclusions}}

In this paper, we have investigated the potential for identifying signals of new
physics at the LHC using a search strategy which focuses on the $\mathrm{jets} + \met$ 
channel and uses $N_j$ and $\met$ as the principal criteria for resolving such signals 
from the SM background.  This approach is particularly suitable in top-rich scenarios,  
which generically give rise to events with large jet multiplicities.
To demonstrate the effectiveness of this search strategy in such scenarios, 
we have examined the detection prospects it affords in an example model whose field
content comprises a gluino $\wtg$, a light, right-handed stop $\st1$, and a
bino-like neutralino $\N1$.  We have shown that for this model, the 
discovery reach provided by this search strategy is comparable to --- and
potentially greater than --- that afforded by other strategies for identifying 
top-rich new physics at the LHC.  These preliminary results provide compelling 
motivation for a more detailed analysis of this discovery reach --- an analysis 
which takes into account more fully the various experimental subtleties 
(detector effects, calibration issues, \etc) involved.

While we have focused on this specific model in order to demonstrate the 
effectiveness of a search strategy focused on $N_j$ and $\met$, we note that 
similar results can be expected for a broad class of top-rich scenarios.   
These include models with additional fermion generations, SUSY scenarios
featuring light stops and heavy first- and second-generation squarks 
(including certain string-theory-inspired scenarios~\cite{LiMaxin9jets1,LiMaxin9jets2}), 
Little Higgs models with T-parity, UED models, and a variety of other new-physics 
scenarios.  Our results serve as motivation for more detailed studies of the discovery 
potential for new physics using such a search strategy in such contexts --- especially 
those in which decay topologies for the heavy fields involve top quarks and invisible
particles almost exclusively.  Indeed, 
in the time since this paper initially appeared, some studies of this 
sort~\cite{RecentLargeMultATLAS} have been undertaken by the ATLAS collaboration.   

The optimal program of event-selection criteria for any given scenario
(and the detection prospects it affords) of course depends
on the production rates for any new strongly-interacting fields involved,
and hence on the masses, spins, and $SU(3)_c$ representations of those fields.
They also depend quite sensitively on the structure of the decay chains
initiated by those new strongly-interacting fields and on the kinematics
of the final-state particles which result from each decay.
For example, the effectiveness of $\met$ as an event-selection criterion 
in any particular scenario depends on how much of the energy released by particle decays
emerges in the form of jets and how much emerges in the form of invisible 
particles.  This balance depends sensitively on the structures and kinematics 
of the decay chains involved, and the more this balance is tilted toward 
invisible particles, the more effective a discriminant a $\met$ cut will 
generally become.  The optimal event-selection criteria depend on other factors
as well.  For example, at $\Lint = 10~\ifb$ and $\sqrt{s} = 7$~TeV, 
we find that there is no advantage to requiring more than eight energetic jets 
or $\met$ significantly in excess of $300$~GeV, since further elevating the 
cuts on these variables tends to reduce the expected number of signal 
events to a negligible level.  However, for higher $\Lint$ or a 
larger $\sqrt{s}$ (as is expected in the coming LHC run), further 
increasing the $N_j$ and $\met$ thresholds cuts beyond those levels may 
indeed prove fruitful.

On a final note, we mention another potentially interesting strategy which could
be useful in identifying signals of new physics in certain top-rich scenarios, and
which can be viewed as complementary to the strategy outlined here.
This is to focus on final states involving bottom quarks plus $\met$ alone.  The
utility of this approach was investigated in Ref.~\cite{AlwallbbPlusMET} in the
context of a model with an additional pair of color-triplet fermions
$T'$ and $B'$, which decay almost exclusively via the channels
$T'\rightarrow t X$ and $B'\rightarrow b X$, where $X$ is a stable
dark-matter particle.  It was found that for $m_X \sim 1$~GeV, the
$pp \rightarrow B' \bar B' \rightarrow b \overline{b} + \met$ and
$pp \rightarrow T' \bar T' \rightarrow t \overline{t} + \met$ channels
offered a comparable reach.  Furthermore, it was argued that improved
sensitivity in the $b \overline{b} + \met$ channel can be expected as $m_X$
increases, provided that the $T'$ is reasonably light, since when this is the
case, the jets produced from top decay tend to be reasonably soft.
This scenario was analyzed in a recent study by the ATLAS collaboration at an integrated
luminosity of $0.83~\ifb$~\cite{ButlerATLAS}.  In this study, the collaboration
searched for evidence of gluino-pair production followed by the decay
$\wtg \rightarrow b\overline{b}\N1$, assuming a branching fraction for this process 
of effectively unity.  For $m_{\widetilde{b}} < 600$~GeV and a neutralino mass set to 
$M_{\N1}=60$~GeV, the results of this study place a bound
$M_{\wtg} \gtrsim 725$~GeV on the gluino mass, which is roughly 
comparable to the expected $5\sigma$ sensitivity the search strategy we have outlined in
this paper for models in which the primary decay channel for the gluino is 
$\wtg \rightarrow t\overline{t}\N1$.  It would be interesting to consider how
the discovery prospects afforded by these two approaches compare in models in which
{\it both} $\wtg \rightarrow b\overline{b}\N1$ and $\wtg \rightarrow t\overline{t}\N1$
channels have sufficient branching fraction --- to wit, models  which involve not 
only top-rich, but also bottom-rich event topologies.

\section{Acknowledgments}

We gratefully acknowledge J.~Alwall, A.~Rajaraman, J.~Rutherfoord,
S.~Su, X.~Tata, and D.~Yaylali for useful discussions.  This work is supported in
part by the Department of Energy under Grant No.~DE-FG02-04ER41291 and by
the National Science Foundation under Grant No.~1066293.  J.~K. would also
like to thank the Aspen Center for Physics for its hospitality during the
period in which this work was being completed.

\end{document}
